\documentclass[prd,nofootinbib,preprint,superscriptaddress]{revtex4}
\pdfoutput=1
\usepackage[T1]{fontenc}
\usepackage{amsmath,amssymb}
\usepackage{epsfig}
\usepackage{graphicx}
\usepackage[usenames,dvipsnames]{color}
\usepackage{subfigure}
\usepackage{slashed}
\usepackage[colorlinks,citecolor=blue]{hyperref}
\usepackage{pdfpages}
\usepackage{color}
\usepackage{comment}

\def\bea{\begin{eqnarray}}
\def\eea{\end{eqnarray}}

\begin{document}


\title{Dark matter from phase transition generated PBH evaporation with gravitational waves signatures}

\author{Debasish Borah}
\email{dborah@iitg.ac.in}
\affiliation{Department of Physics, Indian Institute of Technology Guwahati, Assam 781039, India}

\author{Suruj Jyoti Das}
\email{surujjd@gmail.com}
\affiliation{Particle Theory  and Cosmology Group, Center for Theoretical Physics of the Universe,
Institute for Basic Science (IBS),
 Daejeon, 34126, Korea}

\author{Indrajit Saha}
\email{s.indrajit@iitg.ac.in}
\affiliation{Department of Physics, Indian Institute of Technology Guwahati, Assam 781039, India}

\begin{abstract}
We study the possibility of generating dark matter (DM) purely from ultra-light primordial black hole (PBH) evaporation with the latter being produced from a first order phase transition (FOPT) in the early Universe. If such ultra-light PBH leads to an early matter domination, it can give rise to a doubly peaked gravitational wave (GW) spectrum in Hz-kHz ballpark with the low frequency peak generated from PBH density fluctuations being within near future experimental sensitivity. In the sub-dominant PBH regime, the FOPT generated GW spectrum comes within sensitivity due to absence of entropy dilution. In both the regimes, PBH mass from a few kg can be probed by GW experiments like BBO, ET, CE, UDECIGO etc. while DM mass gets restricted to the superheavy ballpark in the PBH dominance case. Apart from distinct DM mass ranges in the two scenarios, GW observations can differentiate by measuring their distinct spectral shapes.

\end{abstract}

\maketitle

\section{Introduction}
Primordial black holes (PBH), originally proposed in \cite{Zeldovich:1967lct} and later by Hawking \cite{Hawking:1974rv, Hawking:1975vcx} can have several interesting cosmological consequences \cite{Chapline:1975ojl, Carr:1976zz}. A recent review of PBH can be found in \cite{Carr:2020gox}. Depending upon the initial mass $(m_{\rm in})$ and initial energy fraction $(\beta)$, PBH can face different astrophysical and cosmological constraints. Among them, the ultra-light PBH mass window is relatively less constrained as such PBH evaporate by emitting Hawking radiation \cite{Hawking:1974rv, Hawking:1975vcx} before the big bang nucleosynthesis (BBN) epoch. Comparing the evaporation temperature with the BBN temperature $T_\text{ev}> T_\text{BBN}\simeq 4$ MeV leads to the upper bound on this ultra-light PBH mass window. On the other hand, a lower bound can be obtained by using the upper limit on the scale of inflation from cosmic microwave background (CMB) measurements. Using these BBN and CMB bounds simultaneously leads to the allowed initial mass for ultra-light PBH that reads $0.1\,\text{g}\lesssim m_\text{in}\lesssim 3.4\times 10^8\,\text{g}$. As mentioned above, the range of PBH masses in this window is relatively unconstrained~\cite{Carr:2020gox}. It should be noted that the upper bound is obtained only for PBH which evaporates on a timescale smaller than the age of the universe and have a large initial fraction $\beta$ to dominate the energy density at some epoch. Such evaporation of ultra-light PBH can also lead to the production of dark matter (DM) in the universe \cite{Gondolo:2020uqv, Bernal:2020bjf, Green:1999yh, Khlopov:2004tn, Dai:2009hx, Allahverdi:2017sks, Lennon:2017tqq, Hooper:2019gtx, Sandick:2021gew, Fujita:2014hha, Datta:2020bht, JyotiDas:2021shi,Barman:2021ost, Barman:2022gjo, Barman:2022pdo, Cheek:2021odj, Chaudhuri:2023aiv}. As suggested by observations in astrophysics and cosmology based experiments, approximately $26\%$ of the present universe is made up of DM, the particle origin of which is not yet known. If DM has only gravitational interactions as observations indicate, the conventional thermal freeze-out mechanism fails and PBH evaporation provides a very good alternative.

While DM with only gravitational interactions may not have much detection prospects, the ultra-light PBH and their formation mechanisms in the early universe can lead to promising detection prospects.
In spite of this relatively less constrained mass window, it is still possible to probe an ultra-light PBH dominated epoch in the early universe via detection of stochastic gravitational waves (GW) background \footnote{See Ref. \cite{Roshan:2024qnv} for a recent review on stochastic gravitational wave background.}. Interestingly, detection of stochastic GW can also probe the formation mechanism of PBH in the early universe. In the early universe, PBH can be formed in a variety of ways like, from inflationary perturbations \cite{Hawking:1971ei, Carr:1974nx, Wang:2019kaf, Byrnes:2021jka, Braglia:2022phb}, first-order phase transition (FOPT) \cite{Crawford:1982yz, Hawking:1982ga, Moss:1994iq, Kodama:1982sf, Baker:2021nyl, Kawana:2021tde, Huang:2022him, Hashino:2021qoq, Liu:2021svg, Gouttenoire:2023naa, Lewicki:2023ioy, Gehrman:2023qjn, Kim:2023ixo}, the collapse of topological defects \cite{Hawking:1987bn, Deng:2016vzb} etc. Such mechanisms often come up with their own unique GW spectrum. On the other hand, PBH can lead to generation of GW in a variety of ways. The Hawking evaporation of PBH itself can produce gravitons which might constitute an  ultra-high frequency  GW spectrum \cite{Anantua:2008am}. PBH can also lead to GW emission by forming successive mergers \cite{Zagorac:2019ekv}. In addition, the scalar perturbations leading to the formation of PBH can induce GW at second-order \cite{Saito:2008jc}, which can be further enhanced during the PBH evaporation \cite{Inomata:2020lmk}. Finally, the inhomogeneity in the distribution of PBH may also induce GW at second order, as recently studied in \cite{Papanikolaou:2020qtd, Domenech:2020ssp, Domenech:2021wkk}. In the present work, we consider the last possibility which, for ultra-light PBH mass window, can lead to stochastic GW with peak frequencies in the ballpark of ongoing and planned future experiments.

Motivated by this, we study the possibility of simultaneously probing the signatures of formation mechanism and PBH via detection of stochastic GW background with non-trivial implications for dark matter produced solely from PBH evaporation. Recently, FOPT origin of heavy PBH was studied by the authors of \cite{Gehrman:2023esa, Xie:2023cwi, Banerjee:2023brn, Baldes:2023rqv, Banerjee:2023vst, Gouttenoire:2023pxh}. Here, we consider the ultra-light PBH mass window, not considered in earlier works and study the possibility of forming them in FOPT. The combined GW spectrum arising out of the FOPT and density perturbations of ultra-light PBH consists of a unique doubly-peaked feature. For the PBH mass range of interest, the peak in GW spectrum due to FOPT occurs at a frequency higher than the one due to PBH density perturbations. Depending upon the model parameters, the peak frequencies and corresponding blue and red tilted parts of the GW spectrum remain within the sensitivities of several planned experiments like DECIGO \cite{Kawamura:2006up}, BBO\,\cite{Yagi:2011wg}, ET\,\cite{Punturo_2010}, CE\,\cite{LIGOScientific:2016wof}, UDECIGO (UDECIGO-corr) \cite{Sato:2017dkf, Ishikawa:2020hlo} etc. While doubly peaked GW spectrum arises in a scenario with PBH domination, the DM parameter space is typically restricted to the superheavy ballpark. In sub-dominant PBH scenario, DM mass range gets much wider whereas the GW spectrum generated by FOPT comes within the sensitivities of several future experiments. Depending upon the specific scenario, DM masses from GeV to superheavy regime and PBH mass as low as a few kg remains within sensitivities of such GW experiments. It is worth mentioning that in most of the earlier works studying DM production from PBH evaporation, the formation mechanism of the latter is not included. The inclusion of the PBH formation mechanism not only leads to a tighter parameter space but also to complementary detection prospects.

This paper is organised as follows. In section \ref{sec1}, we outline the particle physics setup responsible for FOPT followed by discussion of PBH formation in section \ref{sec2}. In section \ref{sec3} and section \ref{sec4}, we discuss the stochastic GW generated by FOPT and PBH density fluctuations respectively. In section \ref{sec5}, we discuss the combined GW spectrum arising from FOPT and PBH followed by implications for DM phenomenology in section \ref{sec:dm}. We briefly comment on other FOPT origin of PBH not discussed in this work and the corresponding GW signatures in section \ref{pbh:other}. We finally conclude in section \ref{sec6}.

\section{The setup}
\label{sec1}
Among different scenarios relating FOPT to PBH formation \cite{Crawford:1982yz, Hawking:1982ga, Moss:1994iq, Kodama:1982sf, Baker:2021nyl, Kawana:2021tde, Huang:2022him, Hashino:2021qoq, Liu:2021svg, Gouttenoire:2023naa, Lewicki:2023ioy, Gehrman:2023qjn, Kim:2023ixo}, we consider the false vacuum collapse as the origin of PBH formation. In particular, we consider the collapsing Fermi-ball scenario \cite{Kawana:2021tde, Huang:2022him, Marfatia:2021hcp, Tseng:2022jta, Lu:2022jnp, Marfatia:2022jiz, Xie:2023cwi, Gehrman:2023qjn, Kim:2023ixo}. In this scenario, a dark sector fermion $\chi$ with an asymmetry between $\chi$ and $\bar{\chi}$ leads to the formation of Fermi-balls after getting trapped in the false vacuum. Due to the strong Yukawa force among the dark fermions, such Fermi-balls can collapse into PBH. As the dark sector is asymmetric\footnote{Here, we remain agnostic about the origin of such an asymmetry. Example of a UV complete model can be found in \cite{Huang:2022him}. The GW phenomenology discussed here is generic and independent of such UV completions.} and carries a conserved charge, we consider a global dark $U(1)_D$ symmetry which remains unbroken as the Universe goes through a FOPT. The dark sector fermion $\chi$ has a charge $q_\chi$ under this $U(1)_D$ global symmetry while all other fields are neutral. We consider the FOPT to be driven by a singlet scalar $\Phi$. The dark sector fermion $\chi$ is coupled to $\Phi$ through Yukawa interaction. Since fermion and bosons contribute with opposite signs to the effective potential, and fermion $\chi$ has large Yukawa coupling with $\Phi$, additional bosonic degrees of freedom are required to have an effective potential consistent with a strong FOPT while keeping all dimensionless couplings within perturbative limits. We consider an additional singlet scalar $\Phi'$ and doublet scalar $H_2$ for this purpose. However, this choice is arbitrary. For example, if we embed this setup within a new gauge symmetry under which $\Phi$ transforms non-trivially, such extra scalar fields are no longer required. These additional scalars are also neutral under $U(1)_D$ global symmetry and their couplings with $\Phi$ are the only relevant ones for our analysis.

The relevant Lagrangian of these newly introduced fields is given by 
\begin{equation}
    \mathcal{L} \supset -g_{\chi}\Phi \Bar{\chi}\chi-m_{\chi}\Bar{\chi}\chi-V(\Phi,\Phi', H_2)
\end{equation}
where, the scalar potential (suppressing a few allowed terms for simplicity) is
\begin{align}
    V(\Phi,\Phi',H_2) & =\lambda_D \left (|\Phi|^2-\frac{v_D^2}{2} \right)^2 + m_{\Phi'}^2|\Phi'|^2 + \lambda_{\Phi'}|\Phi'|^4 + \lambda_{\Phi\Phi'}|\Phi|^2 |\Phi'|^2 + m_{H_2}^2|H_2|^2 + \lambda_{H_2}|H_2|^4 \nonumber \\
    & + \lambda_{\Phi H_2}|\Phi|^2|H_2|^2 + \lambda_{\Phi H}|\Phi|^2|H|^2.
\end{align}
Here, $v_D$ is the vacuum expectation value (VEV) of the singlet scalar $\Phi ( \equiv \phi/\sqrt{2})$. In order to realize electroweak vacuum, the coupling $\lambda_{\Phi H}$ with the standard model (SM) Higgs $H$ is suppressed . The one-loop correction to the tree-level potential is computed considering the couplings of all the particles to the singlet scalar $\Phi$ and this correction is known as Coleman-Weinberg (CW) potential \cite{Coleman:1973jx}, which can be written as

\begin{equation}
V_{\rm CW}(\phi)=\frac{1}{64 \pi^2}\sum_{i=\phi,\phi',H_2,\chi}n_i M_i^4(\phi)\left \{\log{ \left ( \frac{M_i^2(\phi)}{v^2} \right )}-C_i\right \}
\end{equation}
where, $n_{\Phi}=1$, $n_{\Phi'}=1$, $n_{H_2}=4$ and $n_\chi=2$
and $C_{\Phi,\Phi',H_2,\chi}=\frac{3}{2}$. The physical field-dependent masses of all particles are
\begin{align*}
M_{\Phi}^2(\phi)=\lambda_D(3 \phi^2- v_D^2), \hspace{0.5 cm} M_{\Phi'}^2(\phi)=m_{\Phi'}^2+\frac{\lambda_{\Phi\Phi'}\phi^2}{2}, \\
M_{H_2}^2(\phi)=m_{H_2}^2+\frac{\lambda_{\Phi H_2}\phi^2}{2},  \hspace{0.5 cm} M_\chi^2(\phi)=(m_\chi+g_\chi\phi)^2.  
\end{align*}
Now, the thermal contributions to the effective potential \cite{Dolan:1973qd,Quiros:1999jp} can be expressed as
\begin{equation}
V_T(\phi,T)=\sum_{i=\Phi,\Phi',H_2}\frac{n_iT^4}{2\pi^2}J_B \left(\frac{M_i^2(\phi)}{T^2}\right) - \frac{n_\chi T^4}{2\pi^2}J_F \left(\frac{M_\chi^2(\phi)}{T^2}\right)
\end{equation}
where\\
$$ J_F(x)=\int_{0}^{\infty}dy\, y^2 \log[1+e^{-\sqrt{y^2+x^2}}], \, J_B(x)=\int_{0}^{\infty}dy\, y^2 \log[1-e^{-\sqrt{y^2+x^2}}].$$
Besides this, the Daisy corrections \cite{Fendley:1987ef,Parwani:1991gq,Arnold:1992rz} are also incorporated in thermal contribution to improve perturbative expansion considering Arnold-Espinosa method \cite{Arnold:1992rz} during FOPT such that $V_{\rm thermal}(\phi,T)=V_T(\phi,T) + V_{\rm daisy}(\phi,T)$. The Daisy contribution can be written as
\begin{equation}
V_{\rm daisy}(\phi,T)=-\frac{T}{2\pi^2}\sum_{i=\Phi',H_2} n_i[M_i^3(\phi,T)-M_i^3(\phi)]
\end{equation}
where, $M_i^2(\phi,T)$=$M_i^2(\phi)$ + $\Pi_i(T)$ and the relevant thermal masses are
\begin{align*}
    M_{\Phi'}^2(\phi,T)= m_{\Phi'}^2(\phi)+\left(\frac{\lambda_{\Phi'}}{4}+\frac{\lambda_{\Phi\Phi'}}{6}\right)T^2, \hspace{0.4 cm}
     M_{H_2}^2(\phi,T)= m_{H_2}^2(\phi)+\left(\frac{\lambda_{H_2}}{4}+\frac{\lambda_{\Phi H_2}}{12}\right)T^2.
\end{align*}
So, at finite temperature, the effective potential can be written as 
\begin{equation}
    V_{\rm eff}(\phi,T)=V_{\rm tree}(\phi) + V_{\rm CW}(\phi) + V_{\rm thermal} (\phi,T) .\label{eq:Veff}
\end{equation}
At high temperatures, the Universe lies in a vacuum $\langle \phi \rangle$=0. When the temperature decreases to a critical temperature $T_c$, two degenerate vacua are created which are separated by a barrier. Below this critical temperature, the Universe tunnels through the barrier from false vacuum $\langle \phi \rangle$=0 to true vacuum $\langle \phi \rangle \neq 0 $. The rate of tunneling is estimated by calculating the bounce solution \cite{Linde:1980tt}. The tunneling rate per unit volume is defined in terms of O(3) symmetric bounce action $S_3(T)$ as
\begin{equation}
    \Gamma(T)\sim T^4e^{-S_3(T)/T}.
\end{equation}
 The nucleation temperature $T_n$ is calculated by comparing the tunneling rate with the Hubble expansion rate as
\begin{equation}
    \Gamma(T_n)=\textbf{H}^4(T_n) = \textbf{H}_*^4.
\end{equation}
The volume fraction of false vacuum of the Universe is defined by $p(T) = e^{-\mathcal{I}(T)}$ \cite{Ellis:2018mja, Ellis:2020nnr}, where
\begin{align}
    \mathcal{I}(T) = \frac{4\pi}{3}\int^{T_c}_T \frac{dT'}{T'^4}\frac{\Gamma(T')}{{\bf H}(T')}\left(\int^{T'}_T \frac{d\tilde{T}}{{\bf H}(\tilde{T})}\right)^3.
\end{align}
The percolation temperature $T_p$ is then calculated by using  $\mathcal{I}(T_p) = 0.34$ \cite{Ellis:2018mja} indicating that at least $34\%$ of the comoving volume is occupied by the true vacuum. After the nucleation, an amount of energy is released as latent heat related to the change of energy-momentum tensor across the bubble wall during the FOPT \cite{Caprini:2019egz,Borah:2020wut} given by
\begin{equation}
    \alpha_*=\frac{\epsilon}{\rho_{\rm rad}}
\end{equation}
where, $\epsilon$ =$\left(  \Delta V_{\rm eff} -\frac{T}{4} \frac{\partial \Delta V_{\rm eff} }{\partial T} \right)_{T=T_n}$, $\Delta V_{\rm eff}= V_{\rm eff}(\phi_{\rm false},T)-V_{\rm eff}(\phi_{\rm true},T)$ and $\rho_{\rm rad}=g_*\pi^2T^4/30$. Here, $g_*$ denotes the total relativistic degrees of freedom. The time span of FOPT is denoted by the parameter $\beta/{\textbf{H}(T) }\simeq T\frac{d}{dT} \left(\frac{S_3}{T} \right)$,  whereas, the bubble wall velocity $v_w$, in general, is related to the Jouguet velocity $v_J = \frac{1/\sqrt{3} + \sqrt{\alpha_*^2 + 2\alpha_*/3}}{1+\alpha_*}$. For the type of FOPT we are considering, $v_w \approx v_J$ \cite{Kamionkowski:1993fg, Steinhardt:1981ct, Espinosa:2010hh} 
whereas in the presence of supercooling they are related non-trivially \cite{Lewicki:2021pgr}. In these analyses, the calculation for the action $S_3 (T)$ can be performed numerically by fitting the effective potential given by Eq. \eqref{eq:Veff} to a generic potential \cite{Adams:1993zs}. The details of this procedure is described in Ref. \cite{Borah:2022cdx}.

\section{PBH formation from Fermi-ball}
\label{sec2}
As the FOPT proceeds, the true vacuum expands to cover the whole Universe, while the false vacuum gradually shrinks to a smaller region. At a later stage, the false vacuum region contracted into smaller disconnected volumes, which again split to negligible size at percolation temperature \cite{Ellis:2018mja}. Within these final pockets of false vacuum, some pre-existing asymmetry of $\chi$ fermions can get trapped and the symmetric part annihilates via $\chi \chi\rightarrow$S and $\chi \chi\rightarrow$SS, to final state particles S (either SM particles or $\phi$) through decay or annihilation. Now, the trapped $\chi$'s create a degeneracy pressure. If such a pressure is able to balance the vacuum pressure, these $\chi$ remnants form a bound state which is known as Fermi-ball \cite{Lee:1986tr, Hong:2020est,Kawana:2021tde,DelGrosso:2023trq}. These Fermi-balls formed at $T_*$ defined by $p(T_*)$=0.29. The energy of a Fermi-ball can be written as
\begin{equation}
    E_{\rm FB}=\frac{3\pi}{4}\left (\frac{3}{2\pi} \right)^{2/3}\frac{Q_{\rm FB}^{4/3}}{R} +\frac{4\pi}{3} U_0(T_*) R^3_{\rm FB},
\end{equation}
where, $Q_{\rm FB}=F_\chi^{\rm trap}\frac{n_\chi-n_{\Bar{\chi}}}{n_{\rm FB}^*}$ with $n_X$ denoting number density of $X$, $F_\chi^{\rm trap} \approx 1$ for maximum trapping of $\chi$ inside false vacuum and $U_0$ is vacuum energy. Now, minimizing the Fermi-ball energy, we can get the mass and radius as
\begin{align}
    M_{\rm FB}= Q_{\rm FB}(12\pi^2 U_0(T_*))^{1/4}, \quad \quad R_{\rm FB}^3=\frac{3}{16\pi}\frac{M_{\rm FB}}{U_0(T_*)}.
\end{align}
Within the Fermi-balls, the $\chi$'s  interact through attractive Yukawa interaction $g_\chi\phi\Bar{\chi}\chi$ and the interaction range is $L_\phi(T)= \left(\frac{d^2V_{\rm eff}}{d\phi^2}|_{\phi=0} \right)^{-1/2}$ \cite{Marfatia:2021hcp}. The Yukawa potential energy contribution to the Fermi-ball energy can be expressed as
\begin{equation}
    E_Y\simeq -\frac{3g_\chi^2}{8\pi}\frac{Q_{\rm FB}^2}{R_{\rm FB}} \left(\frac{L_\phi}{R_{\rm FB}}\right)^2.
\end{equation}
At a later stage, when the Yukawa potential energy becomes larger than the Fermi-ball energy, the Fermi-ball becomes unstable and collapses to form PBH \cite{Kawana:2021tde}. We consider that the Fermi-ball formation temperature is close to the nucleation temperature, $T_*\sim T_n $. The initial PBH mass during formation can be estimated in terms of FOPT parameters as \cite{Kawana:2021tde}
\begin{equation}
    M_{\rm in}\sim 1.4\times10^{21}\times v_w^3 \left(\frac{\eta_\chi}{10^{-3}} \right) \left(\frac{100}{g_*}\right)^{1/4}  \left(\frac{100 \hspace{0.1 cm} \rm GeV}{T_*}\right)^{2}  \left(\frac{100}{\beta/\textbf{H}_*}\right)^{3} \alpha_*^{1/4} \hspace{0.2 cm}{\rm g}. \label{eq:mpbh}
\end{equation}
The initial temperature of the black hole is related to its mass as $T_{\rm BH}^{\rm in} \simeq 10^{13} \text{GeV} \left(\frac{1 \text{g}}{M_{\rm in}}\right)$. The initial abundance of PBH which is characterized by the parameter $\beta_{\rm PBH}$, defined as the ratio of the energy density of PBH to the total energy density, is obtained to be \cite{Kawana:2021tde} 
\begin{equation}
    \beta_{\rm PBH} \sim 1.4\times 10^{-15} \times v_w^{-3} \left(\frac{g_*}{100}\right)^{1/2} \left(\frac{T_*}{100 \hspace{0.1 cm} \rm GeV}\right)^3 \left(\frac{\beta/\textbf{H}_*}{100}\right)^3 \left(\frac{M_{\rm in}}{10^{15} \rm g} \right)^{3/2}.\label{eq:betapbh}
\end{equation}
Now, if the initial abundance of PBH is large enough, PBH can dominate the energy density of the Universe over radiation. This puts a lower bound on $\beta_{\rm PBH}$ for PBH domination, given by
\begin{equation}
    \beta_{\rm PBH} \gtrsim \beta_{\text{crit}} \simeq \frac{T_{\rm evap}}{T_n} \,,\label{eq:betacr}
\end{equation}
where $T_{\rm evap} \simeq\left(\frac{9g_{*}(T_{\rm BH}^{\rm in})}{10240}\right)^{\frac{1}{4}}\left(\frac{M_{\rm P}^{5}}{M_{\rm in}^{3}}\right)^{\frac{1}{2}}$ indicates the temperature of the thermal bath when PBH evaporates.  Solving Eq. \eqref{eq:mpbh} and \eqref{eq:betapbh}, we can connect the FOPT parameters $T_n $ and $\beta/\textbf{H}_*$ to the PBH parameters $M_{\rm in}$ and $\beta_{\rm PBH}$ as 
\begin{equation}
    T_n = 1.61 \times 10^{18}~ \beta_{\rm PBH} \left(\frac{100}{g_*}\right)^{1/4} \left(\frac{M_{\rm in}}{1 \hspace{0.1 cm}\rm g}\right)^{-1/2} \left(\frac{\eta_\chi}{10^{-3}}\right)^{-1}\frac{1}{\alpha_*^{1/4}} \hspace{0.1 cm}\text{GeV}\,,\label{eq:Tn}
\end{equation}
\begin{equation}
    \beta/\textbf{H}_* = 1.75 \times 10^{-2} \alpha^{1/4}_* \left(\frac{\eta_\chi}{10^{-3}}\right) v_w \left(\frac{g_*}{100}\right)^{1/12} \beta_{\text{PBH}}^{-2/3}\,.\label{eq:betaH}
\end{equation}
It should be noted that, we require to maintain $g_\chi v_D > T_n$ to ensure that the fermion $\chi$'s get trapped inside the false vacuum and its penetration to the true vacua is kinematically disfavoured.

\section{GW from FOPT}
\label{sec3}
The stochastic GW spectrum from a FOPT can be estimated by considering all the relevant contributions from bubble collisions \cite{Turner:1990rc,Kosowsky:1991ua,Kosowsky:1992rz,Kosowsky:1992vn,Turner:1992tz}, the sound wave \cite{Hindmarsh:2013xza,Giblin:2014qia,Hindmarsh:2015qta,Hindmarsh:2017gnf} and the turbulence \cite{Kamionkowski:1993fg,Kosowsky:2001xp,Caprini:2006jb,Gogoberidze:2007an,Caprini:2009yp,Niksa:2018ofa} of the plasma medium. So, the stochastic GW spectrum can be written as \cite{Caprini:2015zlo}
\begin{align}
    \Omega_{\rm GW}^{\rm PT}(f) &= \Omega_\phi(f) + \Omega_{\rm sw}(f) + \Omega_{\rm turb}(f),
\end{align}
where $\Omega_\phi, \Omega_{\rm sw}, \Omega_{\rm turb}$ are individual contributions from bubble collisions, sound wave of the plasma and turbulence in the plasma respectively. The GW spectrum for bubble collision can be written as  \cite{Caprini:2015zlo}
\begin{equation}
    \Omega_\phi h^2 = 1.67 \times 10^{-5} \left ( \frac{100}{g_*} \right)^{1/3} \left(\frac{{\bf H_*}}{\beta}\right)^2 \left(\frac{\kappa \alpha_*}{1+\alpha_*}\right)^2 \frac{0.11 v^3_w}{0.42+v^2_w} \frac{3.8(f/f_{\rm peak}^{\rm PT, \phi})^{2.8}}{1+2.8 (f/f_{\rm peak}^{\rm PT, \phi})^{3.8}}\,,
\end{equation}
with the peak frequency \cite{Caprini:2015zlo} given by 
\begin{equation}
    f_{\rm peak}^{\rm PT, \phi} = 1.65 \times 10^{-5} {\rm Hz} \left ( \frac{g_*}{100} \right)^{1/6} \left ( \frac{T_n}{100 \; {\rm GeV}} \right ) \frac{0.62}{1.8-0.1v_w+v^2_w} \left(\frac{\beta}{{\bf H_*}}\right).
\end{equation}
The efficiency factor $\kappa_\phi$ for bubble collision can be expressed as \cite{Kamionkowski:1993fg}
\begin{align}
    \kappa_\phi=\frac{1}{1+0.715 \alpha_*}\left(0.715\alpha_* +\frac{4}{27}\sqrt{3\alpha_*/2}\right).
\end{align}
The GW spectrum produced from the sound wave in the plasma can be written as \cite{Caprini:2015zlo,Caprini:2019egz,Guo:2020grp}
\begin{equation}
    \Omega_{\rm sw} h^2 = 2.65 \times 10^{-6} \left ( \frac{100}{g_*} \right)^{1/3} \left(\frac{{\bf H_*}}{\beta}\right) \left(\frac{\kappa_{\rm sw} \alpha_*}{1+\alpha_*}\right)^2 v_w (f/f_{\rm peak}^{\rm PT, sw})^{3} \left(\frac{7}{4+3 (f/f_{\rm peak}^{\rm PT, sw})^{2}} \right)^{7/2} \Upsilon
\end{equation}
and the corresponding peak frequency is given by \cite{Caprini:2015zlo}
\begin{equation}
    f_{\rm peak}^{\rm PT, sw} = 1.65 \times 10^{-5} {\rm Hz} \left ( \frac{g_*}{100} \right)^{1/6} \left ( \frac{T_n}{100 \; {\rm GeV}} \right )  \left(\frac{\beta}{{\bf H_*}}\right) \frac{2}{\sqrt{3}}.
\end{equation}
The efficiency factor for sound wave can be expressed as \cite{Espinosa:2010hh}
\begin{align}
    \kappa_{\rm sw}=\frac{\sqrt{\alpha_*}}{0.135+ \sqrt{0.98+\alpha_*}}.
\end{align}
The suppression factor $\Upsilon=1-\frac{1}{\sqrt{1+2\tau_{sw}H_*}}$ depends on the lifetime of sound wave $\tau_{\rm sw}$\cite{Guo:2020grp} given by $\tau_{\rm sw}\sim R_*/\bar{U}_f$ where mean bubble separation is $R_*=(8\pi)^{1/3}v_w \beta$ and rms fluid velocity is $\bar{U}_f=\sqrt{3\kappa_{sw}\alpha_*/4}$.

Finally, the GW spectrum generated by the turbulence in the plasma is given by \cite{Caprini:2015zlo}
\begin{equation}
    \Omega_{\rm turb} h^2 = 3.35 \times 10^{-4} \left ( \frac{100}{g_*} \right)^{1/3} \left(\frac{{\bf H_*}}{\beta}\right) \left(\frac{\kappa_{\rm turb} \alpha_*}{1+\alpha_*}\right)^{3/2} v_w \frac{(f/f_{\rm peak}^{\rm PT, turb})^{3}}{(1+ f/f_{\rm peak}^{\rm PT, turb})^{11/3} (1+8\pi f/h_*)}
\end{equation}
with the peak frequency as \cite{Caprini:2015zlo}
\begin{equation}
    f_{\rm peak}^{\rm PT, turb} = 1.65 \times 10^{-5} {\rm Hz} \left ( \frac{g_*}{100} \right)^{1/6} \left ( \frac{T_n}{100 \; {\rm GeV}} \right ) \frac{3.5}{2} \left(\frac{\beta}{{\bf H_*}}\right).
\end{equation}
The efficiency factor for turbulence is $\kappa_{\rm turb}\simeq 0.1\kappa_{\rm sw}$ and the inverse Hubble time at the epoch of GW emission, redshifted to today is
\begin{equation}
   h_*= 1.65\times 10^{-5} \frac{T_n}{100 \hspace{0.1 cm} \rm GeV} \left(\frac{g_*}{100}\right)^{1/6}.
\end{equation}
 It is clear from the above expressions that the contribution from sound waves turns out to be the most dominant one and the peak of the total GW spectrum corresponds to the peak frequency of sound waves contribution.
 
\section{GW from PBH density fluctuations}
\label{sec4}

While there are several ways in which PBH can give rise to GW \cite{Anantua:2008am, Zagorac:2019ekv, Saito:2008jc, Hooper:2020evu}, for our purpose of ultralight PBH, the GW spectrum generated by most of these mechanisms lies at very high frequencies. Hence, we rather focus on GW generated because of the inhomogeneity in the PBH distribution after they are formed  \cite{Papanikolaou:2020qtd, Domenech:2020ssp, Domenech:2021wkk, Domenech:2021ztg, Papanikolaou:2022chm}. Such a spectrum is independent of the formation mechanism of PBH, and is within reach of near-future GW detectors.

After PBH are formed, they are inhomogeneously distributed in space following Poisson statistics \cite{Papanikolaou:2020qtd}. These inhomogeneities can induce GW at second order once PBH dominates the energy density of the Universe \cite{Papanikolaou:2020qtd, Domenech:2020ssp}, which gets further enhanced during PBH  evaporation \cite{Domenech:2020ssp}. The dominant contribution to the GW spectrum observed today can be written as \cite{Domenech:2020ssp, Domenech:2021wkk, Borah:2022iym, Barman:2022pdo}
\begin{equation}
    \Omega_{\rm GW}^{\rm PBH}\simeq \Omega_{\rm peak}^{\rm  PBH}\left(\frac{f}{f_{\rm peak}^{\rm PBH}}\right)^{11/3}\Theta
\left(f_{\rm peak}^{\rm PBH}-f\right).\label{eqn:omgw}
\end{equation}
Note that there exists an ultraviolet cutoff that corresponds to length scales smaller than the mean separation between PBH, below which the PBH cannot be treated as a continuous fluid. The associated frequency is estimated to be
\begin{equation}
     f_{\rm peak}^{\rm PBH}\simeq 1.7\times 10^3\,{\rm Hz}\,\left(\frac{M_{\text{in}}}{10^4 \rm g}\right)^{-5/6}.\label{eqn:fpk}
\end{equation}
The peak amplitude, on the other hand, is given by
\begin{equation}
   \Omega_{\rm peak}^{\rm PBH}\simeq 9.67 \times 10^{-11} \left(\frac{k_{\rm UV}}{k_{\rm eva}}\right)^{17/3} \left(\frac{k_{\rm eq}}{k_{\rm UV}}\right)^{8}\,,\label{eq:omgwpbh1}
\end{equation}
where $k_{\rm UV}$ corresponds to the comoving wavenumber corresponding to the mean separation between PBH, whereas $k_{\rm eq}$, $k_{\rm eva}$ indicates the comoving wavenumber of the modes entering the horizon when PBH begins to dominate and when they evaporate respectively. The ratio of the comoving wavenumbers are found using \cite{Domenech:2020ssp} $k_{\rm UV}/ k_*=(\frac{3 M_{\rm in}}{4\pi \rho_{\rm PBH}^{\rm in}})^{-1/3}\bf{H_*}^{-1}$, $k_{\rm eq}/ k_*=\sqrt{2}\beta_{\rm PBH}^{2/3}$ and $k_{\rm eva}/ k_*=(\beta_{\rm PBH}H_{\rm eva}/\bf{H_*})^{1/3}$, where $\rho_{\rm PBH}^{\rm in}$, $k_*$ and $H_{\rm eva}$ ($\bf{H_*}$)  denotes the initial energy density of PBH, wavenumber at the time of PBH formation, Hubble parameter at the time of PBH evaporation (formation) respectively. Expressing explicitly in terms of the PBH formation temperature $T_{*}\simeq T_{n}$, we get
\begin{equation}
    \frac{k_{\rm UV}}{k_{\rm eva}}\simeq 10^6 \left(\frac{M_{\rm in}}{10^4 \rm g}\right)^{2/3},
\end{equation}
\begin{equation}
    \frac{k_{\rm eq}}{k_{\rm UV}}\simeq3.18\times10^{-11} (\beta_{\rm PBH})^{2/3}\left(\frac{M_{\rm in}}{1 ~\rm g}\right)^{1/3} \left(\frac{T_n}{1 ~\rm GeV}\right)^{2/3}.
\end{equation}
Using the above ratio of wavenumbers in Eq. \eqref{eq:omgwpbh1}, we can write the GW amplitude as 
\begin{equation}
     \Omega_{\rm peak}^{\rm PBH}\simeq 4.75\times 10^{-20} \left(\beta_{\rm PBH}\right)^{16/3}\left(\frac{M_{\text{in}}}{10^7 \rm g}\right)^{58/9} \left(\frac{T_n }{100 \hspace{0.1cm}\rm GeV}\right)^{16/3}.\label{eqn:omgpeak}
\end{equation}
Note that for PBH formed through Fermi-ball collapse, $T_n$ in turn depends on the PBH parameters and the asymmetry $\eta_\chi$ through Eq. \eqref{eq:Tn}.

Now, since GW behave as radiation, they can contribute to extra relativistic degrees of freedom during the epoch of BBN. GW from PBH density fluctuations can violate the BBN bound from Planck, if the GW amplitude is very large. This translates into an upper bound on the initial PBH fraction $\beta_{\rm PBH}$, which we find to be given by 
\begin{equation}
     \beta_{\rm PBH} \lesssim \beta_{\rm max} \simeq 1.8 \times 10^{-4} \left(\frac{M_{\rm in}}{10^{4}g}\right)^{-17/32}\left(\frac{\alpha}{0.1}\right)^{3/16}\left(\frac{\beta / {\bf H_*}}{10}\right)^{3/4}\,.\label{eq:betmax}
\end{equation}

\section{The combined spectra and detection prospects}
\label{sec5}
If PBH are formed during the phase transition through Fermi-ball collapse, they can leave their indirect imprints in the GW spectrum. Considering the dominant sound wave contribution and using Eq. \eqref{eq:Tn} and  \eqref{eq:betaH}, we can analytically relate the FOPT GW spectrum to the PBH mass and initial abundance approximately as 
\begin{equation}
    f_{\text{peak}}^{\rm PT}\simeq \frac{5.33 \times 10^{9}}{S^{1/3}} \beta_{\text PBH}^{1/3}\left(\frac{M_{\rm in}}{1~g}\right)^{-1/2} \text{Hz},\label{eq:fpt}
\end{equation}
\begin{equation}
    \Omega_{\rm peak}^{\rm PT} h^2 \simeq \frac{1.42 \times 10^{-7}}{S^{4/3}} \beta_{\text{PBH}}^{2/3} \alpha^{-1/4}_* \eta_{\chi}^{-1} \left(\frac{\kappa_{\rm sw} \alpha_*}{1+\alpha_*}\right)^2.\label{eq:omgpt}
\end{equation}
Here, the factor $S$ accounts for the extra entropy dilution after the gravitational waves from FOPT are produced, which is because of PBH domination and their subsequent evaporation. This makes the frequency and the FOPT generated GW amplitude more red-shifted, compared to the usual scenario without PBH domination. The factor $S$ can be approximated as
\begin{equation}
    S\simeq \frac{T_{\rm dom}}{T_{\rm eva}} = \beta \frac{T_n}{T_{\rm eva}},
\end{equation}
where $T_{\rm dom}$ indicates the temperature of the Universe when PBH starts to dominate. Now, note that the peak frequency depends explicitly both on the PBH mass as well as the initial abundance of PBH, but not on the asymmetry $\eta_{\chi}$. On the other hand, the peak amplitude is not dependent on the PBH mass explicitly, but is sensitive to the initial abundance of PBH  as well as the asymmetry $\eta_{\chi}$.

\begin{figure}
    \centering
    \includegraphics[scale=0.4]{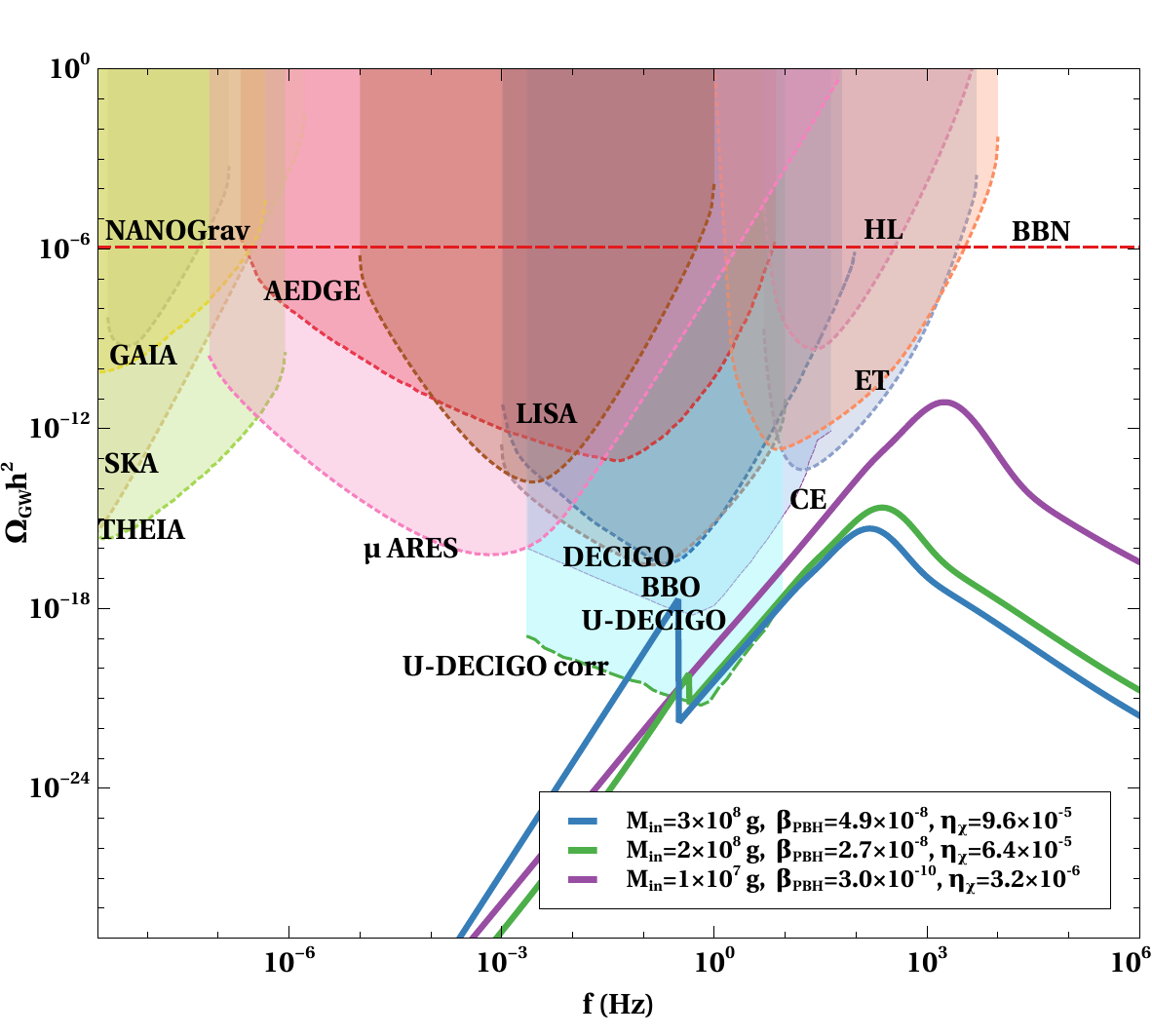}
    \includegraphics[scale=0.4]{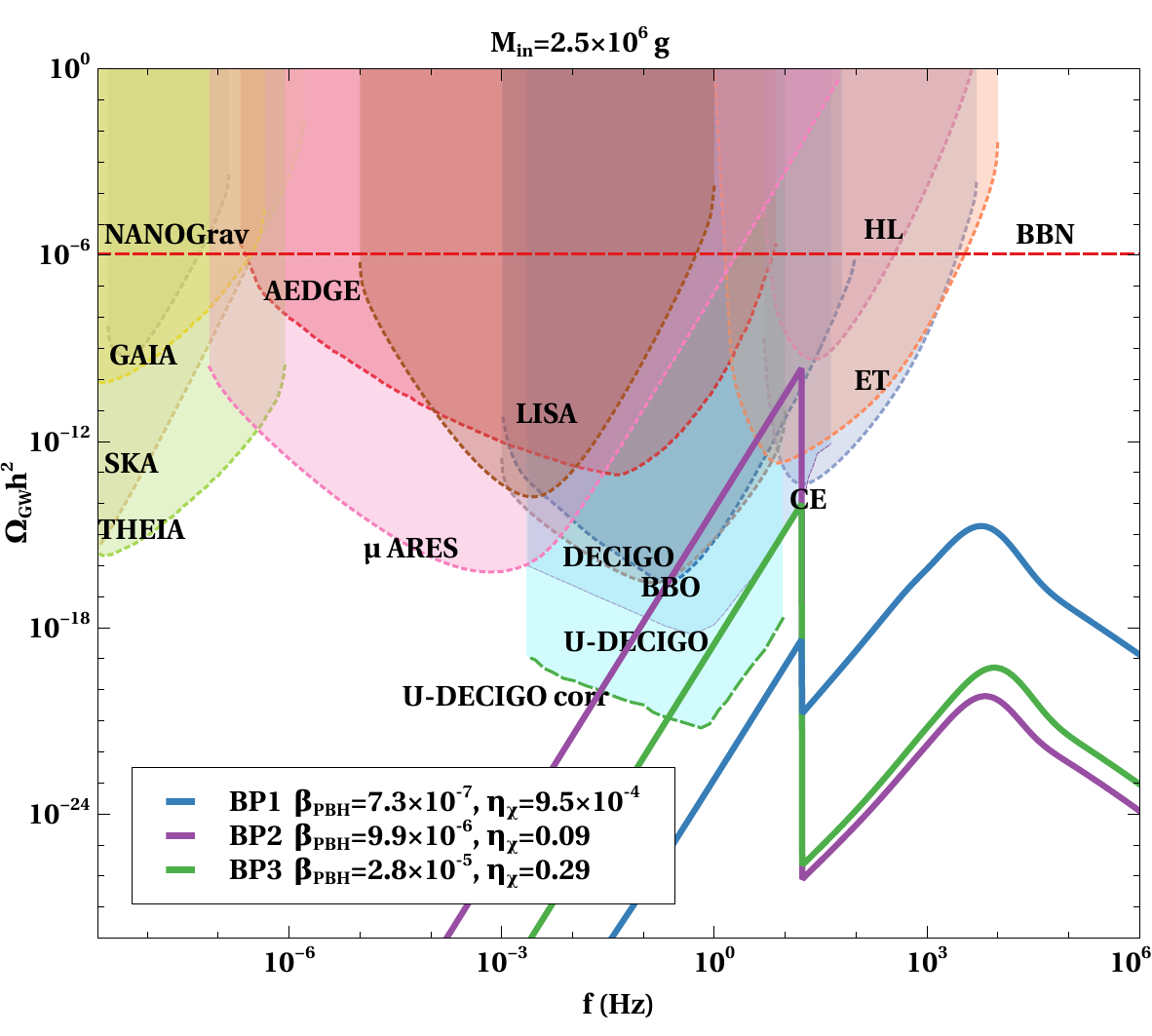}
\caption{Combined GW spectrum from phase transition and PBH density fluctuations for BP4 (see Table \ref{tab1}), considering different values of $M_{\rm in}$ (left panel), and for BP1, BP2 and BP3 for a fixed  $M_{\rm in}$(right panel) but different values of $\beta_{\rm PBH}$. }
    \label{fig:GWspec}
\end{figure}

Considering the contribution of GW from PBH density perturbations, the combined spectrum exhibits a unique doubly peaked feature. While the peak at the higher frequency corresponds to the contribution from FOPT, the sharp peak at the lower frequency is the contribution from PBH density fluctuations. In Fig. \ref{fig:GWspec}, we show the combined GW spectrum varying the initial PBH mass $M_{\rm in}$ (left panel) and initial PBH abundance $\beta_{\rm PBH}$ for a fixed PBH mass (right panel). Note that changing $M_{\rm in}$ also changes $\beta_{\rm PBH}$ in the left panel (cf. Eq. \eqref{eq:betapbh}). The other details of these benchmark points are given in table \ref{tab1}. In all these plots, the experimental sensitivities of SKA \cite{Weltman:2018zrl}, GAIA \cite{Garcia-Bellido:2021zgu}, THEIA \cite{Garcia-Bellido:2021zgu}, $\mu$ARES \cite{Sesana:2019vho}, LISA\,\cite{2017arXiv170200786A}, AEDGE\,\cite{AEDGE:2019nxb}, DECIGO \cite{Kawamura:2006up}, UDECIGO (UDECIGO-corr) \cite{Sato:2017dkf, Ishikawa:2020hlo}, BBO\,\cite{Yagi:2011wg}, ET\,\cite{Punturo_2010}, CE\,\cite{LIGOScientific:2016wof} and aLIGO (HL) \cite{LIGOScientific:2014pky} are shown as shaded regions of different colours. The horizontal dashed red-coloured line corresponds to the BBN limit on effective relativistic degrees of freedom from GW. 

\begin{figure}
    \centering
    \includegraphics[scale=0.45]{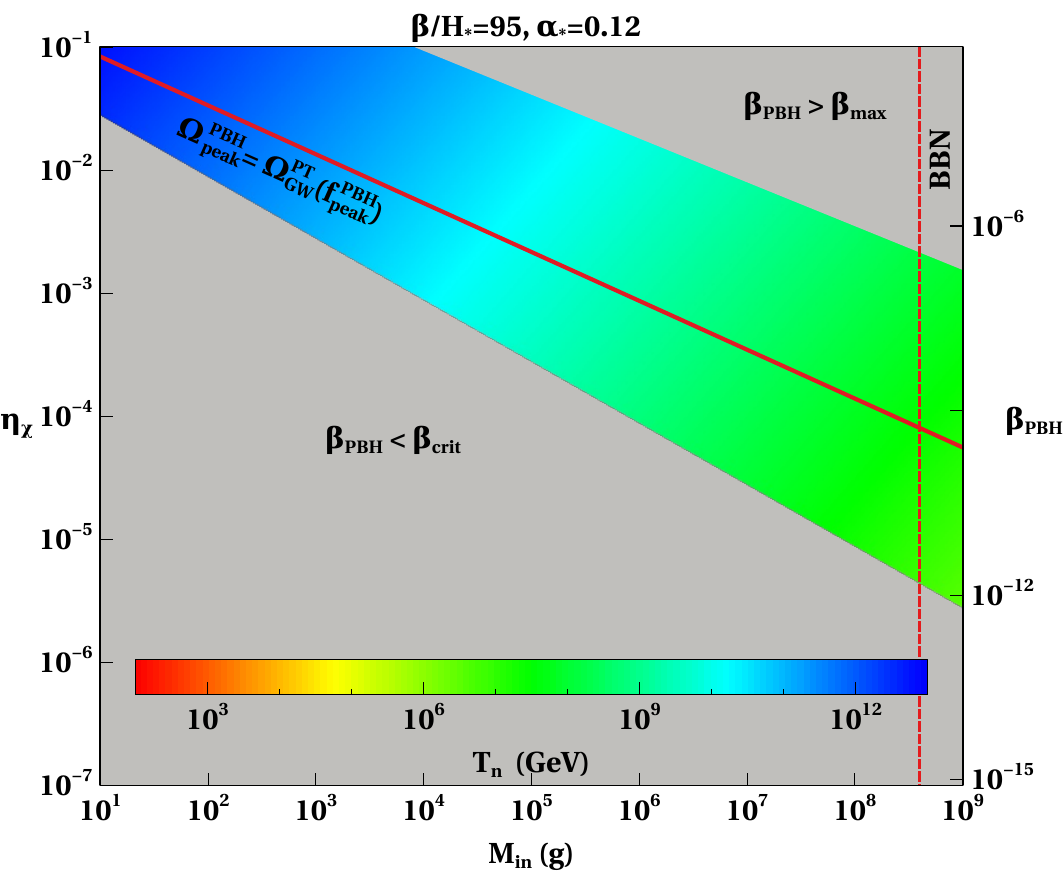}
    \includegraphics[scale=0.45]{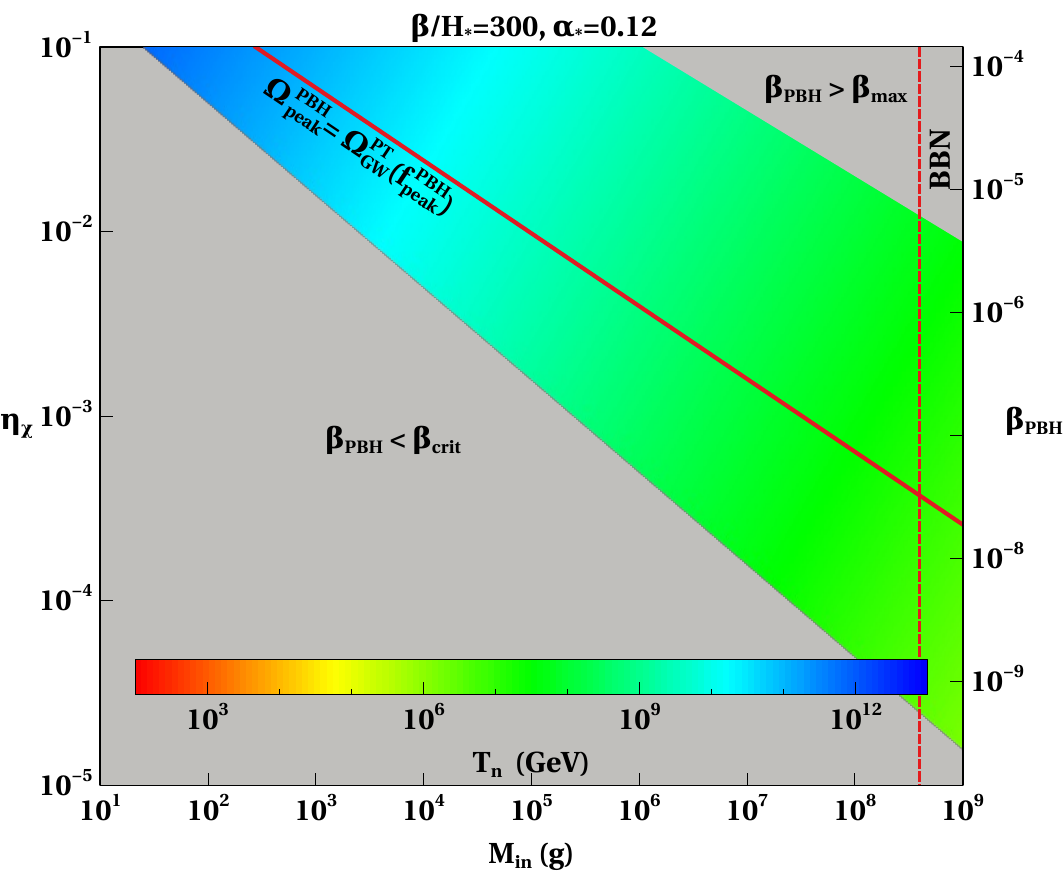}
    \caption{$M_{\rm in}-\eta_{\chi}$ parameter space with $\beta/{\bf H_*}$=95, $\alpha_*$=0.12 (left panel) and $\beta/{\bf H_*}$=300, $\alpha_*$=0.12 (right panel).}
    \label{fig:etaM}
\end{figure}
\begin{figure}
    \centering
   \includegraphics[scale=0.5]{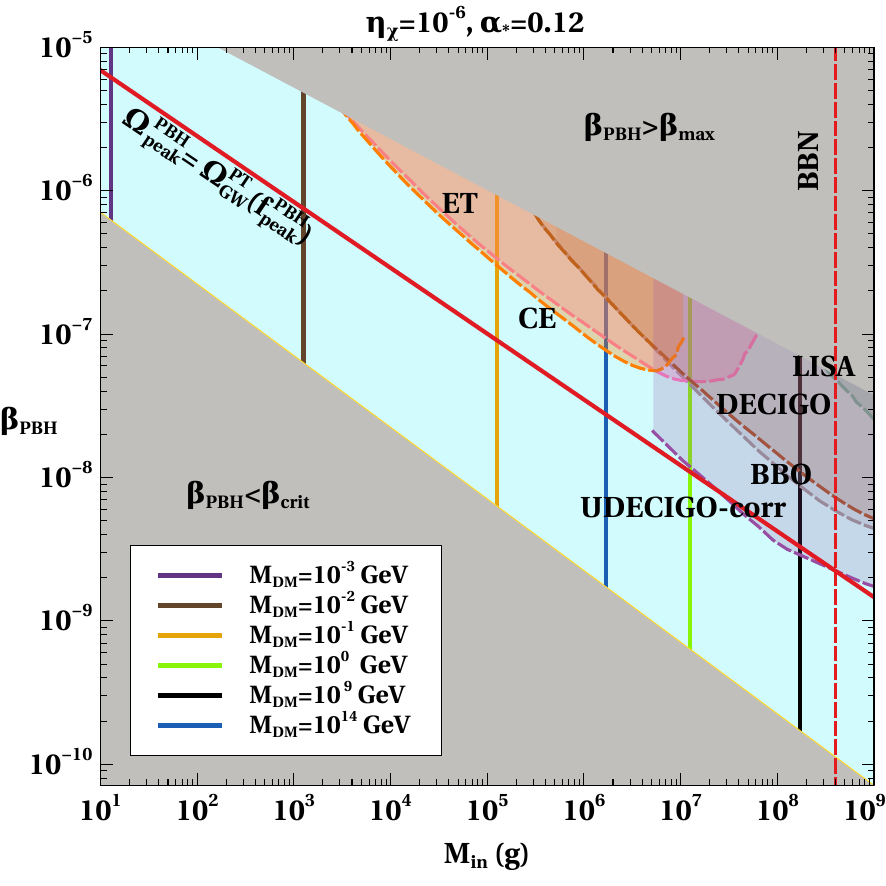}
    \includegraphics[scale=0.5]{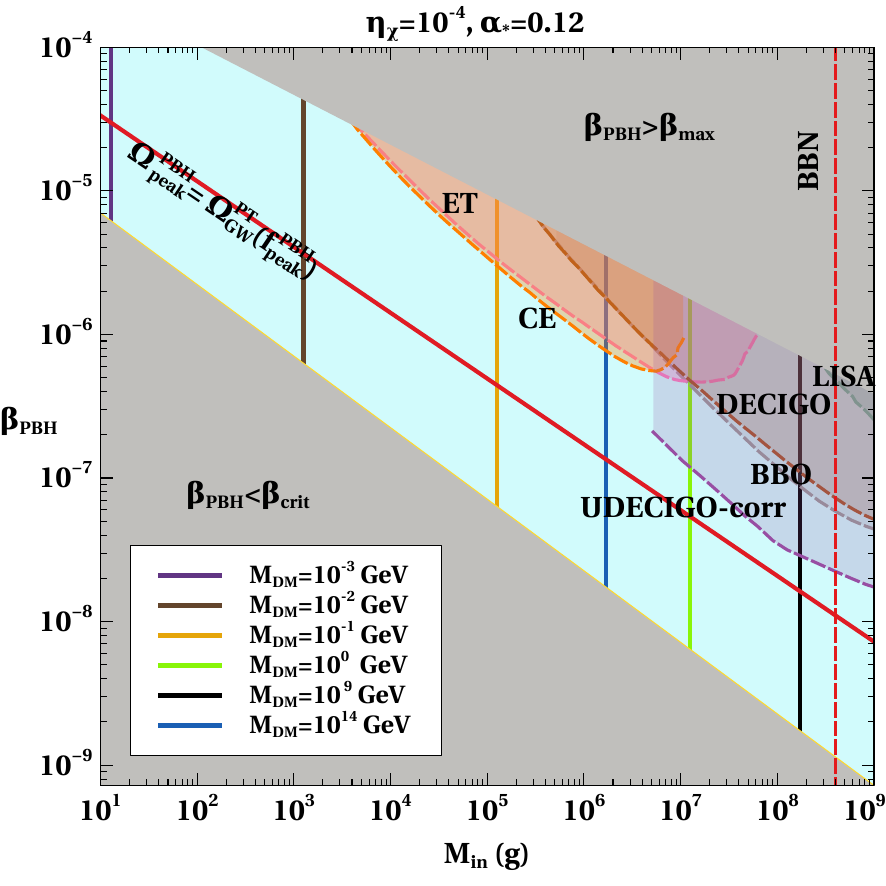}
    \caption{$M_{\rm in}-\beta_{\rm PBH}$ parameter space for $\eta_\chi=10^{-6}$, $\alpha_*$=0.12 (left panel) and $\eta_\chi=10^{-4}$, $\alpha_*$=0.12 (right panel) showing the sensitivity of different experiments to the GW peak from PBH density perturbations.}
    \label{fig:betaM}
\end{figure}

The value of the initial asymmetry $\eta_{\chi}$ depends on the details of the $\chi$ dynamics in the early Universe before the phase transition, and is generally unconstrained. However, the constraints on the PBH abundance $\beta_{\rm PBH}$ can constrain $\eta_{\chi}$, since for PBH formation from Fermi-ball collapse, $\eta_{\chi}$ is related to $\beta_{\rm PBH}$ through Eq. \eqref{eq:Tn}, \eqref{eq:betaH}. In Fig. \ref{fig:etaM}, we show the allowed $\eta_{\chi}$ values in the $M_{\rm PBH}-\eta_{\chi}$ plane, along with corresponding values of the temperature $T_n$ in colour code for two different values of $\beta/{\bf H_*}$. The grey shaded upper triangular region is ruled out since here $\beta_{\rm PBH} > \beta_{\rm max}$, leading to overproduction of GW violating BBN bounds, as discussed earlier. On the other hand, in the grey shaded lower triangular region, $\beta_{\rm PBH}$ turns out to be less than $\beta_{\rm crit}$, which does not lead to PBH domination. In such a case, GW from density fluctuations do not exist, leading to only a single peak feature in the GW spectra, arising from the phase transition. The double peak feature is also absent if the peak of GW contribution from PBH density fluctuation falls below the corresponding GW amplitude from FOPT, i.e. $\Omega_{\rm peak}^{\rm PBH}< \Omega_{GW}^{\rm PT}(f_{\text{peak}}^{\rm PBH})$. As the GW peak arising from the FOPT remains outside the sensitivities of low frequency experiments for most part of the parameter space, we show the parameter space and sensitivities to the GW peak arising from PBH density perturbations in Fig. \ref{fig:betaM}. However, for sub-dominant PBH, the GW peak from FOPT does not suffer any entropy dilution and can be brought within sensitivities of several experiments which we discuss below along with the corresponding implications for DM phenomenology.

\begin{table}[]
    \centering
    \begin{tabular}{|c|c|c|c|c|c|c|c|c|c|c|c|}
    \hline
   & $v_D$ (GeV) & $g_\chi$ & $\lambda_{\Phi\Phi'}$ & $\lambda_{\Phi H_2}$ & $T_c$ (GeV) & $T_n$ (GeV) &$T_p$ (GeV) & $T_*$ (GeV) & $\beta/H_*$ & $\alpha_*$ & $v_J$ \\
    \hline

  BP1 &  $10^{10}$ &  1.4 & 1.8 & 2.5 & $2.8\times 10^9$ & $1.6\times 10^9$ & $1.3\times 10^9$ & $1.3\times 10^9$ & 95 & 0.12  & 0.78\\
    \hline
  BP2 &   $2\times10^9$ &  1.5 & 1.9 & 2.3 & $3.3\times 10^8$ & $2.6\times 10^8$ & $2.6\times 10^8$ & $2.6\times 10^8$ & 1014 & 0.03 & 0.70 \\
     \hline
  BP3 &  $10^9$ &  1.4 & 1.8 & 2.3 & $2.9\times 10^8$ & $2.3\times 10^8$ & $2.2\times 10^8$ & $2.2\times 10^8$ & 1729 & 0.04  & 0.71\\
    \hline
BP4 &  $10^9$ &  1.4 & 1.8 & 2.5 & $2.1\times 10^8$ & $1.2\times 10^8$ & $7.9\times 10^7$ & $7.9\times 10^7$ & 58 & 0.12  & 0.78\\
    \hline
    
    \end{tabular}
    \caption{Benchmark parameters considered in the analysis.}
    \label{tab1}
\end{table}

\begin{figure}
    \centering
   \includegraphics[scale=0.3]{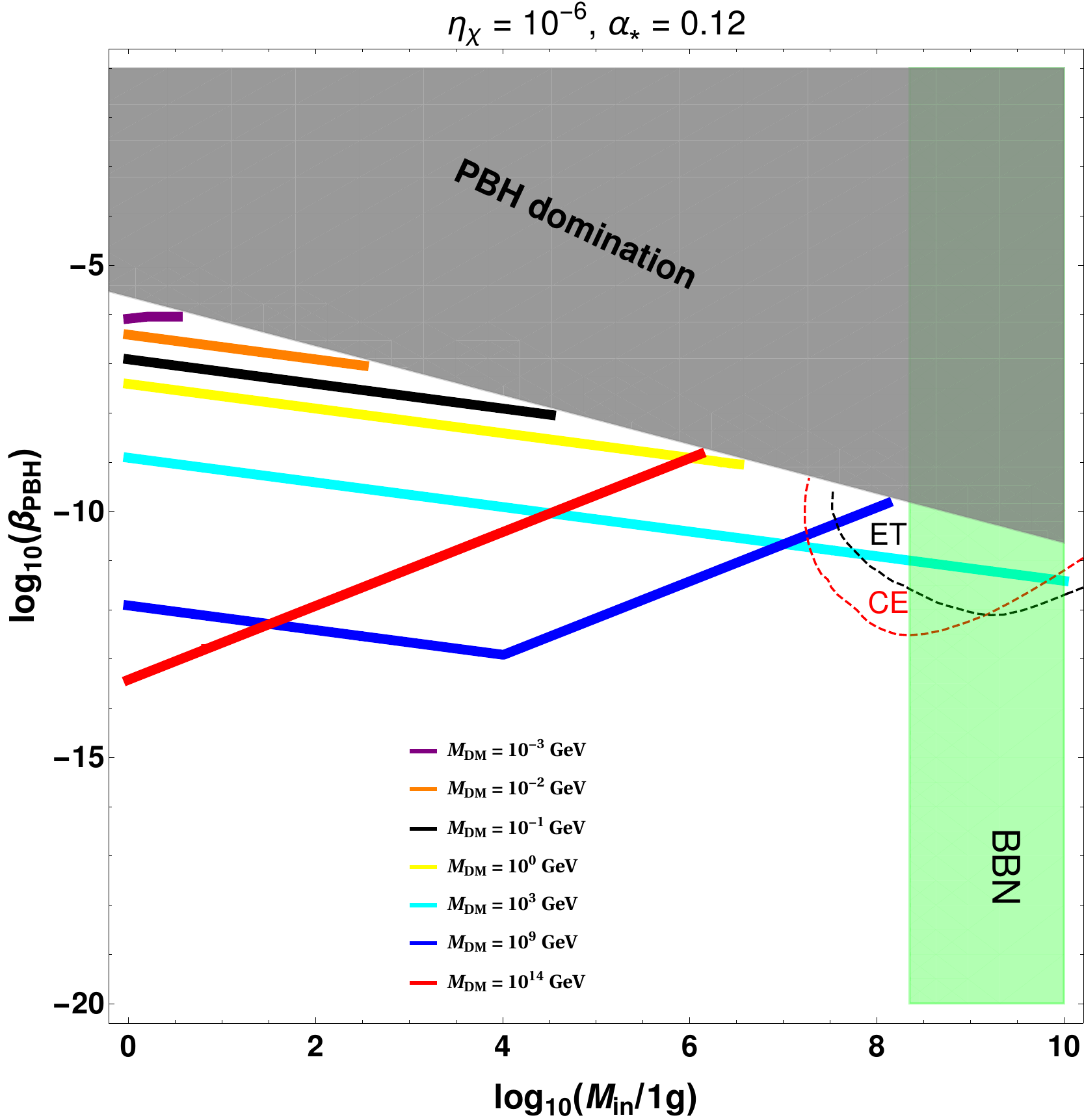}~~
   \includegraphics[scale=0.3]{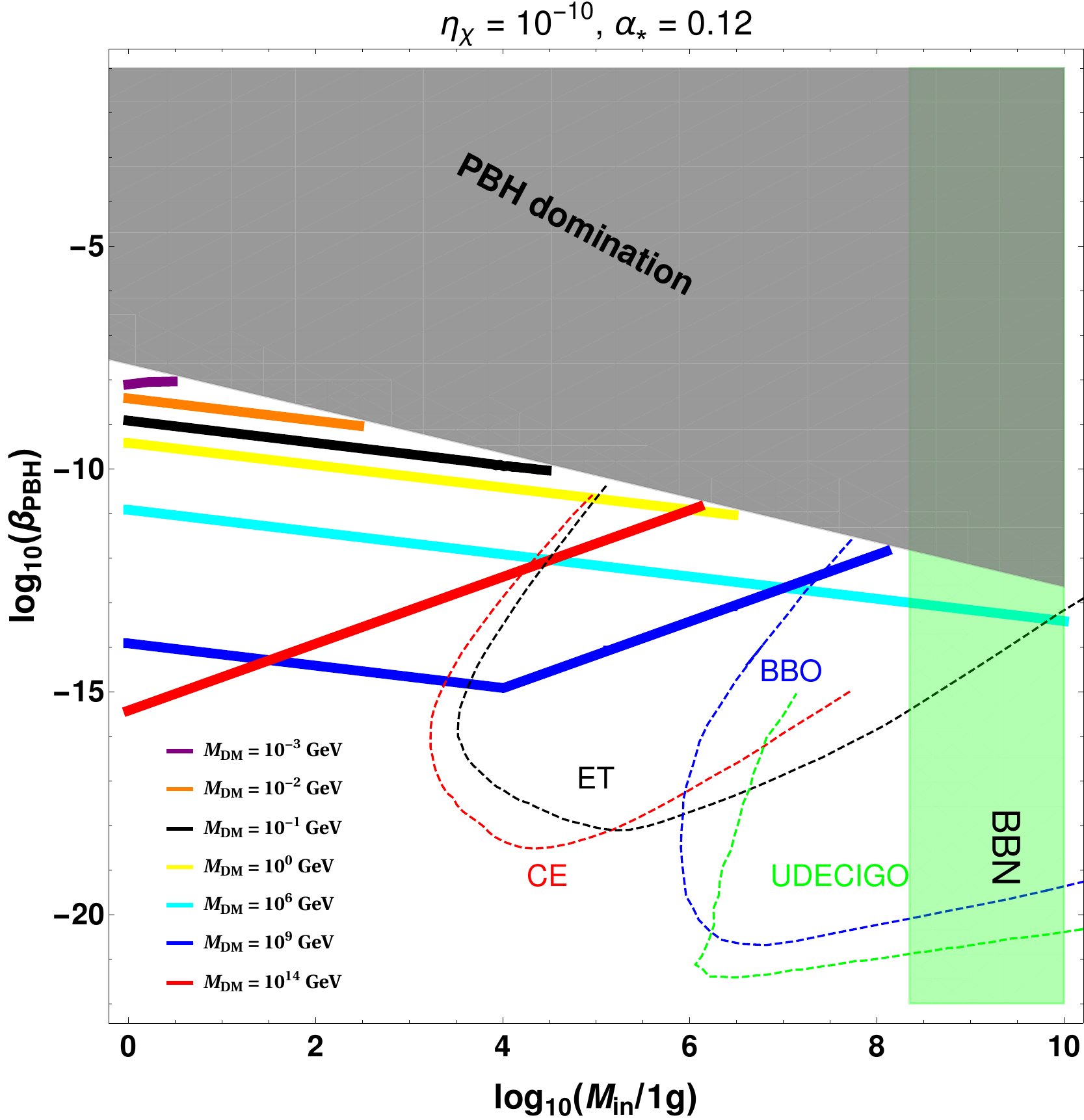}
    \caption{Contours satisfying the observed DM relic abundance for different values of DM mass, considering $\eta_{\chi}= 10^{-6}$ (left panel) and $\eta_{\chi}= 10^{-10}$ (right panel). }
    \label{fig:DM}
\end{figure}

\begin{figure}
    \centering
    \includegraphics[scale=0.4]{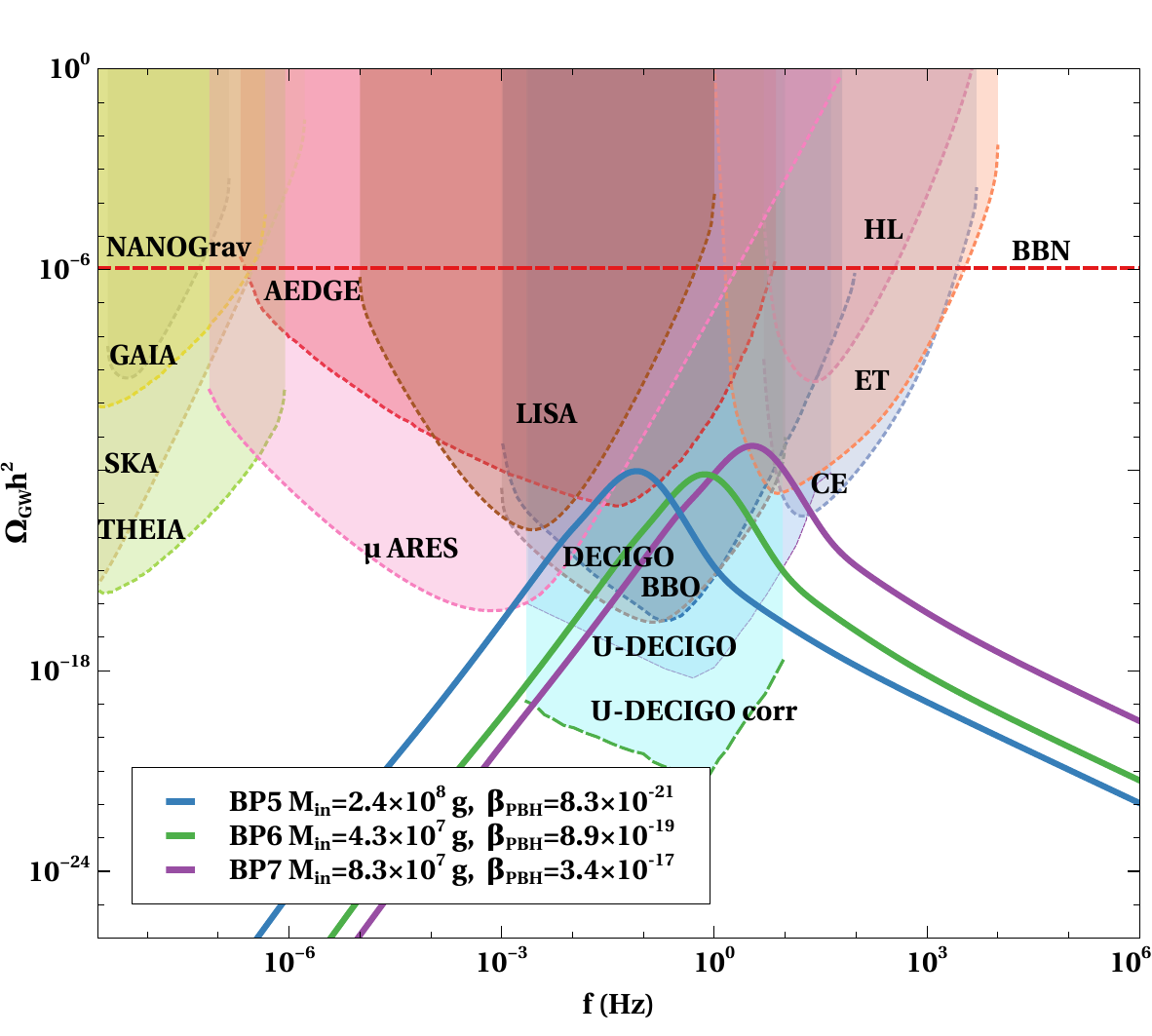}
\caption{GW spectrum from FOPT with sub-dominant PBH for different benchmark points.}
    \label{fig:GWPTspec}
\end{figure}

\begin{figure}
    \centering
   \includegraphics[scale=0.3]{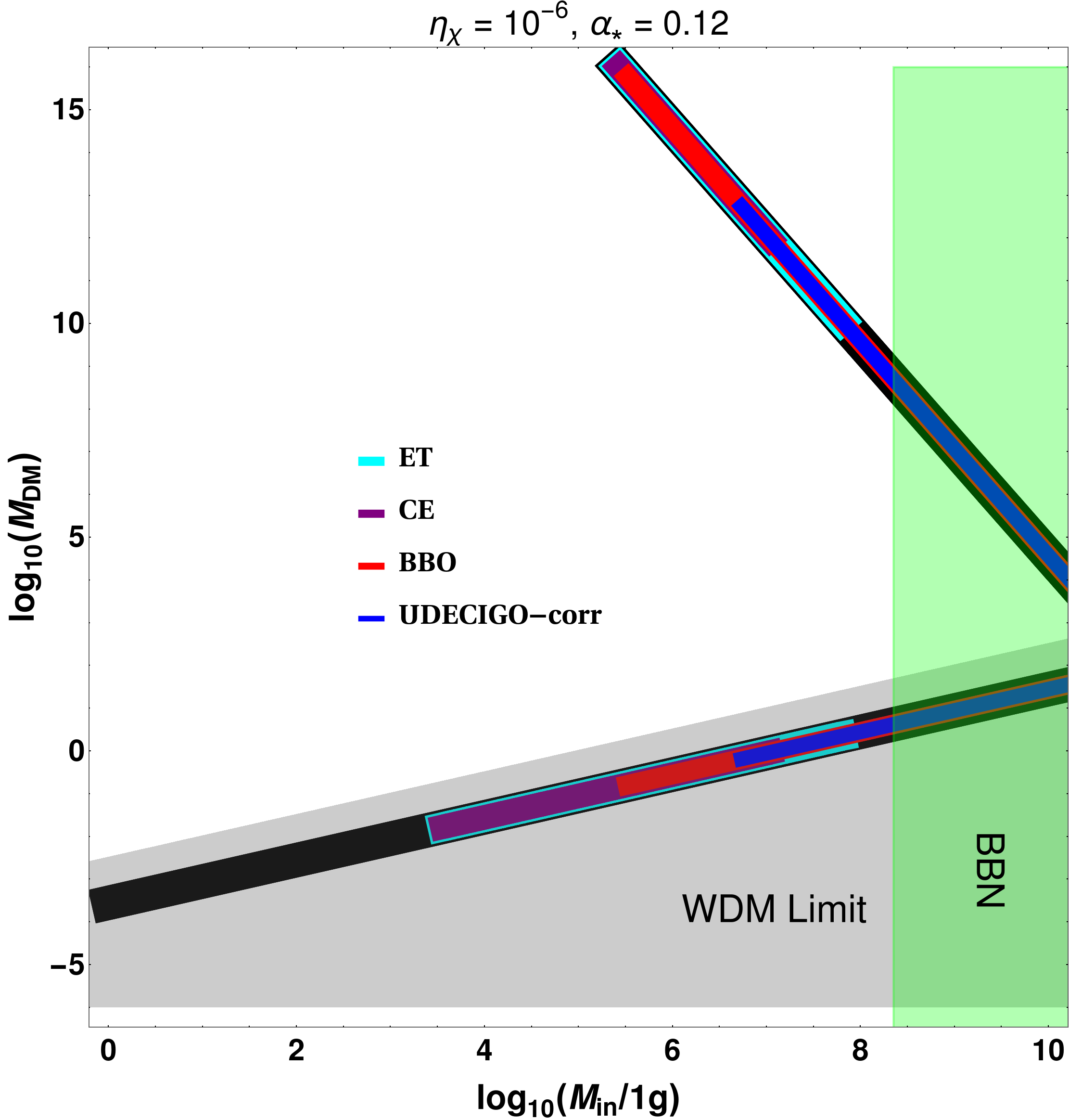}~~
   \includegraphics[scale=0.3]{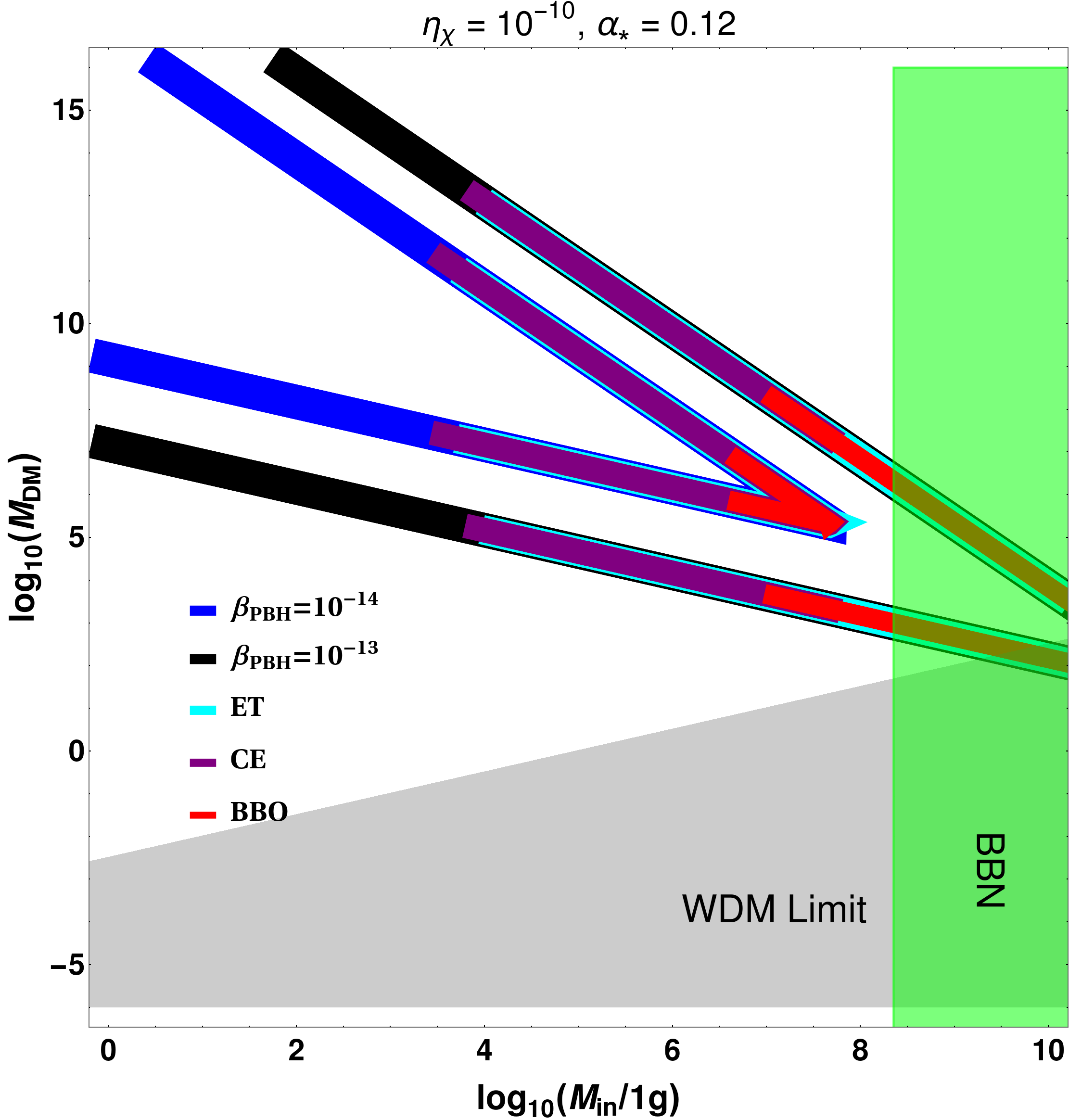}
    \caption{Contours satisfying the observed DM relic abundance in $M_{\rm in}-M_{\rm DM}$ plane, considering $\eta_{\chi}= 10^{-6}$ with PBH domination (left panel) and $\eta_{\chi}= 10^{-10}$ with PBH sub-domination (right panel), along with the future GW experiments that can probe our scenario. The gray-shaded region is constrained from warm dark matter bounds.}
    \label{fig:DMmass}
\end{figure}

\section{Implications for DM phenomenology}
\label{sec:dm}

As seen in the previous section, if PBH dominates the energy density of the Universe we observe a unique doubly-peaked spectrum. However, the second peak at the higher frequency sourced by the FOPT remains out of reach of future GW experiments like ET, CE, DECIGO, BBO. This is primarily because of the extra redshift and dilution arising because of PBH domination and subsequent evaporation (cf. Eq. \eqref{eq:fpt}, \eqref{eq:omgpt}). If PBH do not dominate the energy density, the single peak can be within reach of future GW sensitivity. This can also have interesting consequences in terms of producing the observed DM relic abundance, which otherwise is overproduced in the case of PBH domination, unless superheavy \cite{Samanta:2021mdm, Barman:2022gjo}.

\begin{table}[]
    \centering
    \begin{tabular}{|c|c|c|c|c|c|c|c|c|c|c|c|c|}
    \hline
   & $v_D$ (GeV) & $g_\chi$ & $\lambda_{\Phi\Phi'}$ & $\lambda_{\Phi H_2}$ & $T_c$ (GeV) & $T_n$ (GeV) &$T_p$ (GeV) & $T_*$ (GeV) & $\beta/H_*$ & $\alpha_*$ & $v_J$ & $\eta_\chi$ \\
    \hline
BP5 &  $10^4$ &  1.3 & 1.3 & 2.2 & $2.7\times 10^3$ & $1.6\times 10^3$ & $1.4\times 10^3$ & $1.4\times 10^3$ & 206 & 0.13 & 0.79 & $10^{-12}$ \\
    \hline
BP6 &  $10^5$ &  1.4 & 1.3 & 2 & $3.2\times 10^4$ & $2.1\times 10^4$ & $1.8\times 10^4$ & $1.7\times 10^4$ & 140 & 0.09 & 0.76 & $10^{-11}$ \\
    \hline
BP7 &  $10^6$ &  1.4 & 1.3 & 2.2 & $3.1\times 10^5$ & $1.8\times 10^5$ & $1.2\times 10^5$ & $1.0\times 10^5$ & 78 & 0.12 & 0.78 & $10^{-10}$ \\
    \hline
    
    \end{tabular}
    \caption{Benchmark parameters considered in the analysis for sub-dominant PBH.}
    \label{tab2}
\end{table}

 The dark sector fermion $\chi$ in our setup can not lead to the observed DM abundance, because of being trapped in the false vacuum. However $\chi$ can still be produced from PBH evaporation \cite{Gehrman:2023qjn, Kim:2023ixo} or from the SM bath. Also, some small fraction of $\chi$ can cross from the false vacuum to the true vacuum \cite{Baker:2019ndr}. While a detailed calculation of such trapping and crossing is beyond the scope of our present work, simple conservative estimates, shown in Appendix \ref{appen1} justifies maximum trapping. In all such cases where $\chi$ appears in the true vacuum, the interaction of $\chi$ with the SM bath can play crucial role in deciding its final relic. Since we remain agnostic about the UV completion of the dark sector, we do not consider the possibility of $\chi$ being DM. The amount of $\chi$ produced from the above possible ways can finally decay into the SM bath, before the BBN epoch. As long as $\chi$ is long-lived enough and decay at a temperature $T_{\rm decay} \ll T_*$, the formation temperature of PBH, the phenomenology discussed here remain unaffected. Such a decay of $\chi$ can be introduced via explicit $U(1)_D$ global symmetry breaking higher dimensional operators like $\overline{\ell}_L \tilde{H} \chi \Phi/\Lambda$ where $\ell_L$ denotes the lepton doublet of the SM. Since $\chi$ is lighter in the false vacuum, depending upon the UV completion of the dark sector, it is also possible to ensure DM decay in the true vacuum while preventing the same in the false vacuum due to kinematical restrictions.

If we consider the dark sector to be purely gravitational in agreement with all experimental evidences, it can still be produced from PBH evaporation since the process of Hawking radiation is based solely on gravitational interactions\footnote{PBHs can also leave Planck remnants after evaporation, which can act as DM and have their imprints on GW. See Ref. \cite{Domenech:2023mqk}.}. For the case of sub-dominant PBH ($\beta < \beta_{\rm crit}$), the DM relic density produced from PBH evaporation is found to be\footnote{Here, for our demonstration, we consider the dark sector to consist of scalar particles. Similar results can be obtained for other particles depending on their degrees of freedom.} \cite{Masina:2020xhk, Bernal:2020kse, Bernal:2020bjf, JyotiDas:2021shi}
 \begin{align}
     \Omega_\text{DM}\,h^2  \simeq M_{\rm DM}  \beta_{\rm PBH} \frac{T_n}{M_{\rm in}}
    \begin{cases}
      \left(\frac{M_{\rm in}}{M_P}\right)^2 &\text{for } M_{\rm DM} < T_{\rm BH}^{\rm in}\, , \\[8pt]
       \left(\frac{M_P}{M_{\rm DM}}\right)^2 &\text{for } M_{\rm DM} > T_{\rm BH}^{\rm in}\,,
    \end{cases}\label{eq:rel-dm}\,
 \end{align}
 where $M_{\rm DM}$ indicates the DM mass. Note that $T_{n}$ in turn, depends on the PBH mass and initial fraction (cf. Eq. \eqref{eq:Tn}). On the other hand, in the case of PBH domination, the DM relic density is given by
 \begin{align}
     \Omega_\text{DM}\,h^2  \simeq M_{\rm DM}   \frac{T_{\rm ev}}{M_{\rm in}}
    \begin{cases}
      \left(\frac{M_{\rm in}}{M_P}\right)^2 &\text{for } M_{\rm DM} < T_{\rm BH}^{\rm in}\, , \\[8pt]
       \left(\frac{M_P}{M_{\rm DM}}\right)^2 &\text{for } M_{\rm DM} > T_{\rm BH}^{\rm in}\,.
    \end{cases}\label{eq:rel-dm2}\,
 \end{align}
 As mentioned above, for PBH domination, DM gets over-produced from PBH evaporation irrespective of their spins and only $M_{\rm DM} \gtrsim 10^{9}$ GeV can lead to right abundance for $M_{\rm in} \gtrsim 10^6$ g~\cite{Fujita:2014hha, Samanta:2021mdm, Barman:2022gjo}. DM over-production can also be controlled by choosing sufficiently light DM mass, but it is likely to face constraints from structure formation.

In Fig. \ref{fig:DM}, we show the contours of observed DM abundance in the $M_{\rm in}- \beta_{\rm PBH}$ plane considering two different values of the initial asymmetry $\eta_{\chi}$. As we can see, for PBH domination regime, the DM mass range gets restricted while for sub-dominant PBH scenario, a much wider DM mass range becomes viable. While the value of $\eta_\chi$ does not change the DM parameter space, the GW sensitivity changes as changing $\eta_\chi$ also amounts to changing the FOPT parameters in order to generate the PBH of same initial mass and energy fraction. Clearly, future GW experiments like CE, ET, BBO, UDECIGO can probe a sizeable part of the parameter space in PBH mass ranging from $\mathcal{O}(1000 \, \rm g)$ as well as DM masses ranging from GeV to superheavy ballpark. The detectable GW spectrum in this low frequency regime is purely generated by the FOPT in this case which is also free from any entropy dilution due to PBH evaporation. Fig. \ref{fig:GWPTspec} shows the GW spectrum of a few such benchmark points, the details of which are given in table \ref{tab2}. Similarly, the PBH domination region also remains within GW experimental sensitivities as shown in Fig. \ref{fig:betaM}. While DM is mostly in the superheavy ballpark for such a scenario, PBH mass $M_{\rm in} \gtrsim \mathcal{O}(1000 \, \rm g)$ can be probed by future GW experiments. While PBH mass range within GW sensitivities is similar in both the PBH dominance and sub-dominance cases, it is possible to distinguish these two scenarios at GW experiments due to different GW spectrum arising from PBH density fluctuations (PBH dominance) and FOPT (PBH sub-dominance). 

Finally in Fig. \ref{fig:DMmass}, we show the contours of the observed DM relic density in $M_{\rm in}$- $M_{\rm DM}$ plane, for the two cases, i.e. PBH dominance (left panel)  and subdominance (right panel). For a fixed PBH mass, the DM relic is satisfied at two DM mass values, either when DM is light ($M_{\rm DM} < T_{\rm BH}^{\rm in}$) or when DM is superheavy ($M_{\rm DM} > T_{\rm BH}^{\rm in}$) (cf. Eq. \eqref{eq:rel-dm}, \eqref{eq:rel-dm2}). The free-streaming length of DM is constrained from Lyman-$\alpha$ flux-power spectra~\cite{Irsic:2017ixq, Ballesteros:2020adh, DEramo:2020gpr}, which puts a constraint on light DM depending on the PBH mass \cite{Diamanti:2017xfo, Fujita:2014hha, Masina:2020xhk, Barman:2022gjo}. This is due to the fact that light DM produced from such PBH evaporation can have relativistic speeds more than what is allowed for warm dark matter. This is shown by the gray-shaded region. For the PBH-dominance case, only the superheavy DM regime remains viable, while for subdominant PBH, we can satisfy the correct DM relic density at lighter DM mass as well while being consistent with Lyman-$\alpha$ bounds. The different colored contours along the relic density contour indicate the reach of future GW experiments that can probe the corresponding peak frequencies of the GW spectrum relevant for the PBH and DM mass. The left and the right panels specify the GW peak from density fluctuations and FOPT respectively, both having a one-to-one correspondence with the PBH mass (cf. Eq. \eqref{eqn:fpk}, \eqref{eq:fpt}), considering other parameters to be fixed.      

\section{PBH from other mechanisms related to FOPT}
\label{pbh:other}
Before concluding, we briefly comment on other FOPT origins of PBH and corresponding GW signatures. In other mechanisms without not relying on any dark sector asymmetry or Fermi ball collapse mechanism, PBH are formed when horizon-sized perturbations become more than the critical overdensity $\delta>\delta_c \sim 0.45$ leading to gravitational collapse \cite{Musco:2004ak}. Such type of gravitational collapse as a consequence of FOPT have been studied in several works including \cite{Liu:2021svg,Kawana:2022olo,Hashino:2022tcs, Gouttenoire:2023naa} and references therein. During FOPT, nucleation may get delayed in some Hubble-sized region, and the late decay of the false vacuum increases the total energy density. In an expanding Universe, the radiation energy density keeps on decreasing, whereas in some regions the vacuum energy density remains constant. This leads to total energy density higher than the threshold overdensity and the mass inside such Hubble-sized region collapses to form PBH. After solving the Friedmann equation and radiation energy density evolution equations \cite{Liu:2021svg} in late false vacuum decay region and background region, we can estimate the time, $t_{\rm PBH}$ when PBH formed and the initial PBH mass can be written as $M_{\rm PBH}=\frac{4\pi}{3}{\bf H}^{-3}(t_{\rm PBH})\rho_c$, where $\rho_c$ is critical energy density\footnote{Similar relation is obtained for PBH formed through bubble collisions\cite{Jung:2021mku}.}. In such mechanisms, ultra-light PBH of mass $\sim 8 \times 10^{8}$ g can be formed from FOPT with nucleation temperature of order $10^{13}$ GeV. Thus, if PBH is subdominant, the GW peak from FOPT will be in the ultra-high frequency regime $\sim 10^7$ Hz which is far outside the reach of near-future GW experiments\footnote{There have been several proposed ideas to look for these ultra-high frequency gravitational waves. See Ref. \cite{Aggarwal:2020olq} for a review.}. For PBH domination, the GW peak from FOPT can get redshifted to a lower frequency (depending on $\beta_{\rm PBH}$) because of entropy dilution during PBH evaporation. For instance, considering the PBH mass mentioned above and taking $\beta_{\rm PBH}\sim \beta_{\rm max}$, we find the FOPT peak in the frequency range of order $10^4$ Hz, with a very small amplitude ($\Omega_{\rm peak}^{\rm PT} h^2\lesssim 10^{-25}$) because of the entropy dilution. Moreover, note that a detailed analysis requires calculating the energy fraction PBH $\beta_{PBH}$ through the probability of late false vacuum decay \cite{Kawana:2022olo}. Therefore, the PBH formation mechanism adopted in this work offers more detection prospects irrespective of whether PBH dominates or remains subdominant.

\section{Conclusion}
\label{sec6}
We have studied the detection prospects of an ultra-light primordial black hole formation mechanism in gravitational wave experiments and consequences for production of gravitational dark matter from such PBH evaporation. Considering the Fermi-ball collapse mechanism of PBH in the context of a first-order phase transition, we consider the combined GW spectrum arising from the FOPT as well as PBH density fluctuations with the latter requiring a PBH dominated phase. We show that such a PBH formation mechanism leads to a unique doubly peaked GW spectrum. While the GW spectrum generated from FOPT suffers from entropy dilution at later stages due to ultra-light PBH evaporation, the blue-tilted part of its arm for certain region of parameter space can remain within experimental sensitivity. Combined detection of both the GW peaks require new experiments in kHz frequency ballpark having sufficient sensitivity. The PBH domination scenario mostly produces superheavy DM whereas the sub-dominant PBH can generate a much wider range of DM. While the doubly peaked feature in sub-dominant PBH scenario disappears due to absence of GW generated by PBH density fluctuations, the FOPT generated GW can be brought within sensitivities of future GW experiments like CE, ET, BBO, UDECIGO corresponding to DM mass in GeV to superheavy regime and PBH mass as low as a few kg. The PBH domination regime remains within sensitivities of such experiments for PBH mass as low as a few kg and DM masses mostly in the superheavy ballpark. Due to the strong connection between PBH parameters and FOPT parameters which in turn affect DM mass from relic abundance criteria, we have interesting correlations and complementarity among different observables which can be tested in near future.

\section*{Acknowledgements}
The work of DB is supported by the Science and Engineering Research Board (SERB), Government of India grant MTR/2022/000575 and CRG/2022/000603. The work of SJD was supported by IBS under the project code, IBS-R018-D1. SJD would like to thank the support and hospitality of the Tata Institute of Fundamental Research (TIFR), Mumbai, where a portion of this project was carried out.

\appendix
\section{Particle trapping in false vacuum}
\label{appen1}
Initially, all particles are in the false vacuum state. As the first order phase transition proceeds, particles gain mass in the true vacuum. The bubble of true vacuum expands and tries to capture the particles in the plasma. Due to the dynamical change of mass in the true vacuum, only sufficiently energetic particles are able to enter the bubble while conserving energy. Here, we show a simple quantitative analysis of particles getting trapped or reflected from bubble wall to false vacuum following the works of \cite{Chway:2019kft,Gehrman:2023qjn}. We can estimate the particle flux going to true vacuum in the bubble wall rest frame. The number of particles passing through the bubble wall  along -z direction (considered for simplicity) per area $\Delta S$ in $\Delta t$ time can be written as 
\begin{equation}
    \frac{\Delta N_{in}}{\Delta S}=g_i\int\frac{d^3p}{(2\pi)^3}\int_{r_0}^{r_0-\frac{p_z \Delta t}{p}}dr \mathcal{T}(p) \Theta(-p_z) f(p,x)
    \label{A1}
\end{equation}
where, $r_0$ is bubble radius, $\Theta(-p_z)$ is ensuring particles going inside and $\mathcal{T}(p)=\Theta(-p_z-m_i)$ satisfied energy conservation inside bubble. Also, $g_i$ is degrees of freedom of particle with mass $m_i$ in true vacuum. Here, $f(p,x)$ is the distribution function of particles in false vacuum and it will be Bose-Einstein (BE) or Fermi-Dirac (FD) distribution. The distribution can be approximated to be Maxwell-Boltzmann distribution as $f(p,x)=e^{-\tilde{\gamma} (p+\tilde{v}p_z)/T}$, where $\tilde{v}$ is the velocity of fluid or plasma with respect to bubble wall and $\tilde{\gamma}$ is corresponding Lorentz factor. The flux of particles in bubble wall frame can be calculated from above Eq. \eqref{A1} as
\begin{equation}
    J_w=g_i T^3 \left(\frac{\tilde{\gamma}(1-\tilde{v})m_i/T+1}{4\pi^2 \tilde{\gamma}^3 (1-\tilde{v})^2}  \right) e^{-\frac{\tilde{\gamma}(1-\tilde{v})m_i}{T}}.
\end{equation}
Since, the bubble is expanding with a wall velocity $v_w$, the average number density inside true vacuum in global plasma frame can be expressed as 
\begin{equation}
    n_{\rm in}=\frac{J_w}{\gamma_w v_w}.
\end{equation}
So, the fraction of particle which gets trapped in false vacuum can be written as 
\begin{equation}
    F^{\rm trap}=1-\frac{n_{\rm in}}{n_{\rm eq}}
    \label{ftrap}
\end{equation}
where, $n_{\rm eq}$ is equilibrium number density of particles. In general, the velocity of fluid or plasma $\tilde{v}$ depends on fluid dynamics. From Eq. \eqref{ftrap}, it can be seen that $F^{\rm trap}$ is maximum ($\sim1$) for fluid velocity $\tilde{v}$ O(0.1). 


\begin{thebibliography}{100}

\bibitem{Zeldovich:1967lct}
Y.~B. Zel'dovich and I.~D. Novikov, {\it {The Hypothesis of Cores Retarded
  during Expansion and the Hot Cosmological Model}},  {\em Soviet Astron. AJ
  (Engl. Transl. ),} {\bf 10} (1967) 602.

\bibitem{Hawking:1974rv}
S.~W. Hawking, {\it {Black hole explosions}},  {\em Nature} {\bf 248} (1974)
  30--31.

\bibitem{Hawking:1975vcx}
S.~W. Hawking, {\it {Particle Creation by Black Holes}},  {\em Commun. Math.
  Phys.} {\bf 43} (1975) 199--220. [Erratum: Commun.Math.Phys. 46, 206 (1976)].

\bibitem{Chapline:1975ojl}
G.~F. Chapline, {\it {Cosmological effects of primordial black holes}},  {\em
  Nature} {\bf 253} (1975), no.~5489 251--252.

\bibitem{Carr:1976zz}
B.~J. Carr, {\it {Some cosmological consequences of primordial black-hole
  evaporations}},  {\em Astrophys. J.} {\bf 206} (1976) 8--25.

\bibitem{Carr:2020gox}
B.~Carr, K.~Kohri, Y.~Sendouda, and J.~Yokoyama, {\it {Constraints on
  primordial black holes}},  {\em Rept. Prog. Phys.} {\bf 84} (2021), no.~11
  116902, [\href{http://arxiv.org/abs/2002.12778}{{\tt arXiv:2002.12778}}].

\bibitem{Gondolo:2020uqv}
P.~Gondolo, P.~Sandick, and B.~Shams Es~Haghi, {\it {Effects of primordial
  black holes on dark matter models}},  {\em Phys. Rev. D} {\bf 102} (2020),
  no.~9 095018, [\href{http://arxiv.org/abs/2009.02424}{{\tt
  arXiv:2009.02424}}].

\bibitem{Bernal:2020bjf}
N.~Bernal and O.~Zapata, {\it {Dark Matter in the Time of Primordial Black
  Holes}},  {\em JCAP} {\bf 03} (2021) 015,
  [\href{http://arxiv.org/abs/2011.12306}{{\tt arXiv:2011.12306}}].

\bibitem{Green:1999yh}
A.~M. Green, {\it {Supersymmetry and primordial black hole abundance
  constraints}},  {\em Phys. Rev. D} {\bf 60} (1999) 063516,
  [\href{http://arxiv.org/abs/astro-ph/9903484}{{\tt astro-ph/9903484}}].

\bibitem{Khlopov:2004tn}
M.~Y. Khlopov, A.~Barrau, and J.~Grain, {\it {Gravitino production by
  primordial black hole evaporation and constraints on the inhomogeneity of the
  early universe}},  {\em Class. Quant. Grav.} {\bf 23} (2006) 1875--1882,
  [\href{http://arxiv.org/abs/astro-ph/0406621}{{\tt astro-ph/0406621}}].

\bibitem{Dai:2009hx}
D.-C. Dai, K.~Freese, and D.~Stojkovic, {\it {Constraints on dark matter
  particles charged under a hidden gauge group from primordial black holes}},
  {\em JCAP} {\bf 06} (2009) 023, [\href{http://arxiv.org/abs/0904.3331}{{\tt
  arXiv:0904.3331}}].

\bibitem{Allahverdi:2017sks}
R.~Allahverdi, J.~Dent, and J.~Osinski, {\it {Nonthermal production of dark
  matter from primordial black holes}},  {\em Phys. Rev. D} {\bf 97} (2018),
  no.~5 055013, [\href{http://arxiv.org/abs/1711.10511}{{\tt
  arXiv:1711.10511}}].

\bibitem{Lennon:2017tqq}
O.~Lennon, J.~March-Russell, R.~Petrossian-Byrne, and H.~Tillim, {\it {Black
  Hole Genesis of Dark Matter}},  {\em JCAP} {\bf 04} (2018) 009,
  [\href{http://arxiv.org/abs/1712.07664}{{\tt arXiv:1712.07664}}].

\bibitem{Hooper:2019gtx}
D.~Hooper, G.~Krnjaic, and S.~D. McDermott, {\it {Dark Radiation and Superheavy
  Dark Matter from Black Hole Domination}},  {\em JHEP} {\bf 08} (2019) 001,
  [\href{http://arxiv.org/abs/1905.01301}{{\tt arXiv:1905.01301}}].

\bibitem{Sandick:2021gew}
P.~Sandick, B.~S. Es~Haghi, and K.~Sinha, {\it {Asymmetric reheating by
  primordial black holes}},  {\em Phys. Rev. D} {\bf 104} (2021), no.~8 083523,
  [\href{http://arxiv.org/abs/2108.08329}{{\tt arXiv:2108.08329}}].

\bibitem{Fujita:2014hha}
T.~Fujita, M.~Kawasaki, K.~Harigaya, and R.~Matsuda, {\it {Baryon asymmetry,
  dark matter, and density perturbation from primordial black holes}},  {\em
  Phys. Rev. D} {\bf 89} (2014), no.~10 103501,
  [\href{http://arxiv.org/abs/1401.1909}{{\tt arXiv:1401.1909}}].

\bibitem{Datta:2020bht}
S.~Datta, A.~Ghosal, and R.~Samanta, {\it {Baryogenesis from ultralight
  primordial black holes and strong gravitational waves from cosmic strings}},
  {\em JCAP} {\bf 08} (2021) 021, [\href{http://arxiv.org/abs/2012.14981}{{\tt
  arXiv:2012.14981}}].

\bibitem{JyotiDas:2021shi}
S.~Jyoti~Das, D.~Mahanta, and D.~Borah, {\it {Low scale leptogenesis and dark
  matter in the presence of primordial black holes}},
  \href{http://arxiv.org/abs/2104.14496}{{\tt arXiv:2104.14496}}.

\bibitem{Barman:2021ost}
B.~Barman, D.~Borah, S.~J. Das, and R.~Roshan, {\it {Non-thermal origin of
  asymmetric dark matter from inflaton and primordial black holes}},  {\em
  JCAP} {\bf 03} (2022), no.~03 031,
  [\href{http://arxiv.org/abs/2111.08034}{{\tt arXiv:2111.08034}}].

\bibitem{Barman:2022gjo}
B.~Barman, D.~Borah, S.~Das~Jyoti, and R.~Roshan, {\it {Cogenesis of Baryon
  Asymmetry and Gravitational Dark Matter from PBH}},
  \href{http://arxiv.org/abs/2204.10339}{{\tt arXiv:2204.10339}}.

\bibitem{Barman:2022pdo}
B.~Barman, D.~Borah, S.~Jyoti~Das, and R.~Roshan, {\it {Gravitational wave
  signatures of a PBH-generated baryon-dark matter coincidence}},  {\em Phys.
  Rev. D} {\bf 107} (2023), no.~9 095002,
  [\href{http://arxiv.org/abs/2212.00052}{{\tt arXiv:2212.00052}}].

\bibitem{Cheek:2021odj}
A.~Cheek, L.~Heurtier, Y.~F. Perez-Gonzalez, and J.~Turner, {\it {Primordial
  Black Hole Evaporation and Dark Matter Production: I. Solely Hawking
  radiation}},  \href{http://arxiv.org/abs/2107.00013}{{\tt arXiv:2107.00013}}.

\bibitem{Chaudhuri:2023aiv}
A.~Chaudhuri, B.~Coleppa, and K.~Loho, {\it {Dark matter production from two
  evaporating PBH distributions}},  {\em Phys. Rev. D} {\bf 108} (2023), no.~3
  035040, [\href{http://arxiv.org/abs/2301.08588}{{\tt arXiv:2301.08588}}].

\bibitem{Roshan:2024qnv}
R.~Roshan and G.~White, {\it {Using gravitational waves to see the first second
  of the Universe}},  \href{http://arxiv.org/abs/2401.04388}{{\tt
  arXiv:2401.04388}}.

\bibitem{Hawking:1971ei}
S.~Hawking, {\it {Gravitationally collapsed objects of very low mass}},  {\em
  Mon. Not. Roy. Astron. Soc.} {\bf 152} (1971) 75.

\bibitem{Carr:1974nx}
B.~J. Carr and S.~W. Hawking, {\it {Black holes in the early Universe}},  {\em
  Mon. Not. Roy. Astron. Soc.} {\bf 168} (1974) 399--415.

\bibitem{Wang:2019kaf}
S.~Wang, T.~Terada, and K.~Kohri, {\it {Prospective constraints on the
  primordial black hole abundance from the stochastic gravitational-wave
  backgrounds produced by coalescing events and curvature perturbations}},
  {\em Phys. Rev. D} {\bf 99} (2019), no.~10 103531,
  [\href{http://arxiv.org/abs/1903.05924}{{\tt arXiv:1903.05924}}]. [Erratum:
  Phys.Rev.D 101, 069901 (2020)].

\bibitem{Byrnes:2021jka}
C.~T. Byrnes and P.~S. Cole, {\it {Lecture notes on inflation and primordial
  black holes}},  12, 2021.
\newblock \href{http://arxiv.org/abs/2112.05716}{{\tt arXiv:2112.05716}}.

\bibitem{Braglia:2022phb}
M.~Braglia, A.~Linde, R.~Kallosh, and F.~Finelli, {\it {Hybrid
  $\alpha$-attractors, primordial black holes and gravitational wave
  backgrounds}},  \href{http://arxiv.org/abs/2211.14262}{{\tt
  arXiv:2211.14262}}.

\bibitem{Crawford:1982yz}
M.~Crawford and D.~N. Schramm, {\it {Spontaneous Generation of Density
  Perturbations in the Early Universe}},  {\em Nature} {\bf 298} (1982)
  538--540.

\bibitem{Hawking:1982ga}
S.~W. Hawking, I.~G. Moss, and J.~M. Stewart, {\it {Bubble Collisions in the
  Very Early Universe}},  {\em Phys. Rev. D} {\bf 26} (1982) 2681.

\bibitem{Moss:1994iq}
I.~G. Moss, {\it {Singularity formation from colliding bubbles}},  {\em Phys.
  Rev. D} {\bf 50} (1994) 676--681.

\bibitem{Kodama:1982sf}
H.~Kodama, M.~Sasaki, and K.~Sato, {\it {Abundance of Primordial Holes Produced
  by Cosmological First Order Phase Transition}},  {\em Prog. Theor. Phys.}
  {\bf 68} (1982) 1979.

\bibitem{Baker:2021nyl}
M.~J. Baker, M.~Breitbach, J.~Kopp, and L.~Mittnacht, {\it {Primordial Black
  Holes from First-Order Cosmological Phase Transitions}},
  \href{http://arxiv.org/abs/2105.07481}{{\tt arXiv:2105.07481}}.

\bibitem{Kawana:2021tde}
K.~Kawana and K.-P. Xie, {\it {Primordial black holes from a cosmic phase
  transition: The collapse of Fermi-balls}},  {\em Phys. Lett. B} {\bf 824}
  (2022) 136791, [\href{http://arxiv.org/abs/2106.00111}{{\tt
  arXiv:2106.00111}}].

\bibitem{Huang:2022him}
P.~Huang and K.-P. Xie, {\it {Primordial black holes from an electroweak phase
  transition}},  {\em Phys. Rev. D} {\bf 105} (2022), no.~11 115033,
  [\href{http://arxiv.org/abs/2201.07243}{{\tt arXiv:2201.07243}}].

\bibitem{Hashino:2021qoq}
K.~Hashino, S.~Kanemura, and T.~Takahashi, {\it {Primordial black holes as a
  probe of strongly first-order electroweak phase transition}},  {\em Phys.
  Lett. B} {\bf 833} (2022) 137261,
  [\href{http://arxiv.org/abs/2111.13099}{{\tt arXiv:2111.13099}}].

\bibitem{Liu:2021svg}
J.~Liu, L.~Bian, R.-G. Cai, Z.-K. Guo, and S.-J. Wang, {\it {Primordial black
  hole production during first-order phase transitions}},  {\em Phys. Rev. D}
  {\bf 105} (2022), no.~2 L021303, [\href{http://arxiv.org/abs/2106.05637}{{\tt
  arXiv:2106.05637}}].

\bibitem{Gouttenoire:2023naa}
Y.~Gouttenoire and T.~Volansky, {\it {Primordial Black Holes from Supercooled
  Phase Transitions}},  \href{http://arxiv.org/abs/2305.04942}{{\tt
  arXiv:2305.04942}}.

\bibitem{Lewicki:2023ioy}
M.~Lewicki, P.~Toczek, and V.~Vaskonen, {\it {Primordial black holes from
  strong first-order phase transitions}},  {\em JHEP} {\bf 09} (2023) 092,
  [\href{http://arxiv.org/abs/2305.04924}{{\tt arXiv:2305.04924}}].

\bibitem{Gehrman:2023qjn}
T.~C. Gehrman, B.~Shams Es~Haghi, K.~Sinha, and T.~Xu, {\it {Recycled Dark
  Matter}},  \href{http://arxiv.org/abs/2310.08526}{{\tt arXiv:2310.08526}}.

\bibitem{Kim:2023ixo}
T.~Kim, P.~Lu, D.~Marfatia, and V.~Takhistov, {\it {Regurgitated Dark Matter}},
   \href{http://arxiv.org/abs/2309.05703}{{\tt arXiv:2309.05703}}.

\bibitem{Hawking:1987bn}
S.~W. Hawking, {\it {Black Holes From Cosmic Strings}},  {\em Phys. Lett. B}
  {\bf 231} (1989) 237--239.

\bibitem{Deng:2016vzb}
H.~Deng, J.~Garriga, and A.~Vilenkin, {\it {Primordial black hole and wormhole
  formation by domain walls}},  {\em JCAP} {\bf 04} (2017) 050,
  [\href{http://arxiv.org/abs/1612.03753}{{\tt arXiv:1612.03753}}].

\bibitem{Anantua:2008am}
R.~Anantua, R.~Easther, and J.~T. Giblin, {\it {GUT-Scale Primordial Black
  Holes: Consequences and Constraints}},  {\em Phys. Rev. Lett.} {\bf 103}
  (2009) 111303, [\href{http://arxiv.org/abs/0812.0825}{{\tt
  arXiv:0812.0825}}].

\bibitem{Zagorac:2019ekv}
J.~L. Zagorac, R.~Easther, and N.~Padmanabhan, {\it {GUT-Scale Primordial Black
  Holes: Mergers and Gravitational Waves}},  {\em JCAP} {\bf 06} (2019) 052,
  [\href{http://arxiv.org/abs/1903.05053}{{\tt arXiv:1903.05053}}].

\bibitem{Saito:2008jc}
R.~Saito and J.~Yokoyama, {\it {Gravitational wave background as a probe of the
  primordial black hole abundance}},  {\em Phys. Rev. Lett.} {\bf 102} (2009)
  161101, [\href{http://arxiv.org/abs/0812.4339}{{\tt arXiv:0812.4339}}].
  [Erratum: Phys.Rev.Lett. 107, 069901 (2011)].

\bibitem{Inomata:2020lmk}
K.~Inomata, M.~Kawasaki, K.~Mukaida, T.~Terada, and T.~T. Yanagida, {\it
  {Gravitational Wave Production right after a Primordial Black Hole
  Evaporation}},  {\em Phys. Rev. D} {\bf 101} (2020), no.~12 123533,
  [\href{http://arxiv.org/abs/2003.10455}{{\tt arXiv:2003.10455}}].

\bibitem{Papanikolaou:2020qtd}
T.~Papanikolaou, V.~Vennin, and D.~Langlois, {\it {Gravitational waves from a
  universe filled with primordial black holes}},  {\em JCAP} {\bf 03} (2021)
  053, [\href{http://arxiv.org/abs/2010.11573}{{\tt arXiv:2010.11573}}].

\bibitem{Domenech:2020ssp}
G.~Dom\`enech, C.~Lin, and M.~Sasaki, {\it {Gravitational wave constraints on
  the primordial black hole dominated early universe}},  {\em JCAP} {\bf 04}
  (2021) 062, [\href{http://arxiv.org/abs/2012.08151}{{\tt arXiv:2012.08151}}].
  [Erratum: JCAP 11, E01 (2021)].

\bibitem{Domenech:2021wkk}
G.~Dom\`enech, V.~Takhistov, and M.~Sasaki, {\it {Exploring evaporating
  primordial black holes with gravitational waves}},  {\em Phys. Lett. B} {\bf
  823} (2021) 136722, [\href{http://arxiv.org/abs/2105.06816}{{\tt
  arXiv:2105.06816}}].

\bibitem{Gehrman:2023esa}
T.~C. Gehrman, B.~Shams Es~Haghi, K.~Sinha, and T.~Xu, {\it {The primordial
  black holes that disappeared: connections to dark matter and MHz-GHz
  gravitational Waves}},  {\em JCAP} {\bf 10} (2023) 001,
  [\href{http://arxiv.org/abs/2304.09194}{{\tt arXiv:2304.09194}}].

\bibitem{Xie:2023cwi}
K.-P. Xie, {\it {Pinning down the primordial black hole formation mechanism
  with gamma-rays and gravitational waves}},  {\em JCAP} {\bf 06} (2023) 008,
  [\href{http://arxiv.org/abs/2301.02352}{{\tt arXiv:2301.02352}}].

\bibitem{Banerjee:2023brn}
I.~K. Banerjee and U.~K. Dey, {\it {Probing the origin of primordial black
  holes through novel gravitational wave spectrum}},  {\em JCAP} {\bf 07}
  (2023) 024, [\href{http://arxiv.org/abs/2305.07569}{{\tt arXiv:2305.07569}}].

\bibitem{Baldes:2023rqv}
I.~Baldes and M.~O. Olea-Romacho, {\it {Primordial black holes as dark matter:
  Interferometric tests of phase transition origin}},
  \href{http://arxiv.org/abs/2307.11639}{{\tt arXiv:2307.11639}}.

\bibitem{Banerjee:2023vst}
I.~K. Banerjee and U.~K. Dey, {\it {Gravitational Waves from Superradiance to
  Probe the Origin of Primordial Black Holes}},
  \href{http://arxiv.org/abs/2311.02876}{{\tt arXiv:2311.02876}}.

\bibitem{Gouttenoire:2023pxh}
Y.~Gouttenoire, {\it {Primordial Black Holes from Conformal Higgs}},
  \href{http://arxiv.org/abs/2311.13640}{{\tt arXiv:2311.13640}}.

\bibitem{Kawamura:2006up}
S.~Kawamura et~al., {\it {The Japanese space gravitational wave antenna
  DECIGO}},  {\em Class. Quant. Grav.} {\bf 23} (2006) S125--S132.

\bibitem{Yagi:2011wg}
K.~Yagi and N.~Seto, {\it {Detector configuration of DECIGO/BBO and
  identification of cosmological neutron-star binaries}},  {\em Phys. Rev. D}
  {\bf 83} (2011) 044011, [\href{http://arxiv.org/abs/1101.3940}{{\tt
  arXiv:1101.3940}}]. [Erratum: Phys.Rev.D 95, 109901 (2017)].

\bibitem{Punturo_2010}
{\bf ET Collaboration} Collaboration, M.~Punturo~et al, {\it The einstein
  telescope: a third-generation gravitational wave observatory},  {\em
  Classical and Quantum Gravity} {\bf 27} (sep, 2010) 194002.

\bibitem{LIGOScientific:2016wof}
{\bf LIGO Scientific} Collaboration, B.~P. Abbott et~al., {\it {Exploring the
  Sensitivity of Next Generation Gravitational Wave Detectors}},  {\em Class.
  Quant. Grav.} {\bf 34} (2017), no.~4 044001,
  [\href{http://arxiv.org/abs/1607.08697}{{\tt arXiv:1607.08697}}].

\bibitem{Sato:2017dkf}
S.~Sato et~al., {\it {The status of DECIGO}},  {\em J. Phys. Conf. Ser.} {\bf
  840} (2017), no.~1 012010.

\bibitem{Ishikawa:2020hlo}
T.~Ishikawa et~al., {\it {Improvement of the target sensitivity in DECIGO by
  optimizing its parameters for quantum noise including the effect of
  diffraction loss}},  {\em Galaxies} {\bf 9} (2021), no.~1 14,
  [\href{http://arxiv.org/abs/2012.11859}{{\tt arXiv:2012.11859}}].

\bibitem{Marfatia:2021hcp}
D.~Marfatia and P.-Y. Tseng, {\it {Correlated signals of first-order phase
  transitions and primordial black hole evaporation}},  {\em JHEP} {\bf 08}
  (2022) 001, [\href{http://arxiv.org/abs/2112.14588}{{\tt arXiv:2112.14588}}].
  [Erratum: JHEP 08, 249 (2022)].

\bibitem{Tseng:2022jta}
P.-Y. Tseng and Y.-M. Yeh, {\it {511 keV line and primordial black holes from
  first-order phase transitions}},  {\em JCAP} {\bf 08} (2023) 035,
  [\href{http://arxiv.org/abs/2209.01552}{{\tt arXiv:2209.01552}}].

\bibitem{Lu:2022jnp}
P.~Lu, K.~Kawana, and A.~Kusenko, {\it {Late-forming primordial black holes:
  Beyond the CMB era}},  {\em Phys. Rev. D} {\bf 107} (2023), no.~10 103037,
  [\href{http://arxiv.org/abs/2210.16462}{{\tt arXiv:2210.16462}}].

\bibitem{Marfatia:2022jiz}
D.~Marfatia and P.-Y. Tseng, {\it {Boosted dark matter from primordial black
  holes produced in a first-order phase transition}},  {\em JHEP} {\bf 04}
  (2023) 006, [\href{http://arxiv.org/abs/2212.13035}{{\tt arXiv:2212.13035}}].

\bibitem{Coleman:1973jx}
S.~R. Coleman and E.~J. Weinberg, {\it {Radiative Corrections as the Origin of
  Spontaneous Symmetry Breaking}},  {\em Phys. Rev. D} {\bf 7} (1973)
  1888--1910.

\bibitem{Dolan:1973qd}
L.~Dolan and R.~Jackiw, {\it {Symmetry Behavior at Finite Temperature}},  {\em
  Phys. Rev. D} {\bf 9} (1974) 3320--3341.

\bibitem{Quiros:1999jp}
M.~Quiros, {\it {Finite temperature field theory and phase transitions}},  in
  {\em {ICTP Summer School in High-Energy Physics and Cosmology}},
  pp.~187--259, 1, 1999.
\newblock \href{http://arxiv.org/abs/hep-ph/9901312}{{\tt hep-ph/9901312}}.

\bibitem{Fendley:1987ef}
P.~Fendley, {\it {The Effective Potential and the Coupling Constant at High
  Temperature}},  {\em Phys. Lett. B} {\bf 196} (1987) 175--180.

\bibitem{Parwani:1991gq}
R.~R. Parwani, {\it {Resummation in a hot scalar field theory}},  {\em Phys.
  Rev. D} {\bf 45} (1992) 4695,
  [\href{http://arxiv.org/abs/hep-ph/9204216}{{\tt hep-ph/9204216}}]. [Erratum:
  Phys.Rev.D 48, 5965 (1993)].

\bibitem{Arnold:1992rz}
P.~B. Arnold and O.~Espinosa, {\it {The Effective potential and first order
  phase transitions: Beyond leading-order}},  {\em Phys. Rev. D} {\bf 47}
  (1993) 3546, [\href{http://arxiv.org/abs/hep-ph/9212235}{{\tt
  hep-ph/9212235}}]. [Erratum: Phys.Rev.D 50, 6662 (1994)].

\bibitem{Linde:1980tt}
A.~D. Linde, {\it {Fate of the False Vacuum at Finite Temperature: Theory and
  Applications}},  {\em Phys. Lett. B} {\bf 100} (1981) 37--40.

\bibitem{Ellis:2018mja}
J.~Ellis, M.~Lewicki, and J.~M. No, {\it {On the Maximal Strength of a
  First-Order Electroweak Phase Transition and its Gravitational Wave Signal}},
   {\em JCAP} {\bf 04} (2019) 003, [\href{http://arxiv.org/abs/1809.08242}{{\tt
  arXiv:1809.08242}}].

\bibitem{Ellis:2020nnr}
J.~Ellis, M.~Lewicki, and V.~Vaskonen, {\it {Updated predictions for
  gravitational waves produced in a strongly supercooled phase transition}},
  {\em JCAP} {\bf 11} (2020) 020, [\href{http://arxiv.org/abs/2007.15586}{{\tt
  arXiv:2007.15586}}].

\bibitem{Caprini:2019egz}
C.~Caprini et~al., {\it {Detecting gravitational waves from cosmological phase
  transitions with LISA: an update}},  {\em JCAP} {\bf 03} (2020) 024,
  [\href{http://arxiv.org/abs/1910.13125}{{\tt arXiv:1910.13125}}].

\bibitem{Borah:2020wut}
D.~Borah, A.~Dasgupta, K.~Fujikura, S.~K. Kang, and D.~Mahanta, {\it
  {Observable Gravitational Waves in Minimal Scotogenic Model}},  {\em JCAP}
  {\bf 08} (2020) 046, [\href{http://arxiv.org/abs/2003.02276}{{\tt
  arXiv:2003.02276}}].

\bibitem{Kamionkowski:1993fg}
M.~Kamionkowski, A.~Kosowsky, and M.~S. Turner, {\it {Gravitational radiation
  from first order phase transitions}},  {\em Phys. Rev. D} {\bf 49} (1994)
  2837--2851, [\href{http://arxiv.org/abs/astro-ph/9310044}{{\tt
  astro-ph/9310044}}].

\bibitem{Steinhardt:1981ct}
P.~J. Steinhardt, {\it {Relativistic Detonation Waves and Bubble Growth in
  False Vacuum Decay}},  {\em Phys. Rev. D} {\bf 25} (1982) 2074.

\bibitem{Espinosa:2010hh}
J.~R. Espinosa, T.~Konstandin, J.~M. No, and G.~Servant, {\it {Energy Budget of
  Cosmological First-order Phase Transitions}},  {\em JCAP} {\bf 06} (2010)
  028, [\href{http://arxiv.org/abs/1004.4187}{{\tt arXiv:1004.4187}}].

\bibitem{Lewicki:2021pgr}
M.~Lewicki, M.~Merchand, and M.~Zych, {\it {Electroweak bubble wall expansion:
  gravitational waves and baryogenesis in Standard Model-like thermal plasma}},
   {\em JHEP} {\bf 02} (2022) 017, [\href{http://arxiv.org/abs/2111.02393}{{\tt
  arXiv:2111.02393}}].

\bibitem{Adams:1993zs}
F.~C. Adams, {\it {General solutions for tunneling of scalar fields with
  quartic potentials}},  {\em Phys. Rev. D} {\bf 48} (1993) 2800--2805,
  [\href{http://arxiv.org/abs/hep-ph/9302321}{{\tt hep-ph/9302321}}].

\bibitem{Borah:2022cdx}
D.~Borah, A.~Dasgupta, and I.~Saha, {\it {Leptogenesis and dark matter through
  relativistic bubble walls with observable gravitational waves}},  {\em JHEP}
  {\bf 11} (2022) 136, [\href{http://arxiv.org/abs/2207.14226}{{\tt
  arXiv:2207.14226}}].

\bibitem{Lee:1986tr}
T.~D. Lee and Y.~Pang, {\it {Fermion Soliton Stars and Black Holes}},  {\em
  Phys. Rev. D} {\bf 35} (1987) 3678.

\bibitem{Hong:2020est}
J.-P. Hong, S.~Jung, and K.-P. Xie, {\it {Fermi-ball dark matter from a
  first-order phase transition}},  {\em Phys. Rev. D} {\bf 102} (2020), no.~7
  075028, [\href{http://arxiv.org/abs/2008.04430}{{\tt arXiv:2008.04430}}].

\bibitem{DelGrosso:2023trq}
L.~Del~Grosso, G.~Franciolini, P.~Pani, and A.~Urbano, {\it {Fermion soliton
  stars}},  {\em Phys. Rev. D} {\bf 108} (2023), no.~4 044024,
  [\href{http://arxiv.org/abs/2301.08709}{{\tt arXiv:2301.08709}}].

\bibitem{Turner:1990rc}
M.~S. Turner and F.~Wilczek, {\it {Relic gravitational waves and extended
  inflation}},  {\em Phys. Rev. Lett.} {\bf 65} (1990) 3080--3083.

\bibitem{Kosowsky:1991ua}
A.~Kosowsky, M.~S. Turner, and R.~Watkins, {\it {Gravitational radiation from
  colliding vacuum bubbles}},  {\em Phys. Rev. D} {\bf 45} (1992) 4514--4535.

\bibitem{Kosowsky:1992rz}
A.~Kosowsky, M.~S. Turner, and R.~Watkins, {\it {Gravitational waves from first
  order cosmological phase transitions}},  {\em Phys. Rev. Lett.} {\bf 69}
  (1992) 2026--2029.

\bibitem{Kosowsky:1992vn}
A.~Kosowsky and M.~S. Turner, {\it {Gravitational radiation from colliding
  vacuum bubbles: envelope approximation to many bubble collisions}},  {\em
  Phys. Rev. D} {\bf 47} (1993) 4372--4391,
  [\href{http://arxiv.org/abs/astro-ph/9211004}{{\tt astro-ph/9211004}}].

\bibitem{Turner:1992tz}
M.~S. Turner, E.~J. Weinberg, and L.~M. Widrow, {\it {Bubble nucleation in
  first order inflation and other cosmological phase transitions}},  {\em Phys.
  Rev. D} {\bf 46} (1992) 2384--2403.

\bibitem{Hindmarsh:2013xza}
M.~Hindmarsh, S.~J. Huber, K.~Rummukainen, and D.~J. Weir, {\it {Gravitational
  waves from the sound of a first order phase transition}},  {\em Phys. Rev.
  Lett.} {\bf 112} (2014) 041301, [\href{http://arxiv.org/abs/1304.2433}{{\tt
  arXiv:1304.2433}}].

\bibitem{Giblin:2014qia}
J.~T. Giblin and J.~B. Mertens, {\it {Gravitional radiation from first-order
  phase transitions in the presence of a fluid}},  {\em Phys. Rev. D} {\bf 90}
  (2014), no.~2 023532, [\href{http://arxiv.org/abs/1405.4005}{{\tt
  arXiv:1405.4005}}].

\bibitem{Hindmarsh:2015qta}
M.~Hindmarsh, S.~J. Huber, K.~Rummukainen, and D.~J. Weir, {\it {Numerical
  simulations of acoustically generated gravitational waves at a first order
  phase transition}},  {\em Phys. Rev. D} {\bf 92} (2015), no.~12 123009,
  [\href{http://arxiv.org/abs/1504.03291}{{\tt arXiv:1504.03291}}].

\bibitem{Hindmarsh:2017gnf}
M.~Hindmarsh, S.~J. Huber, K.~Rummukainen, and D.~J. Weir, {\it {Shape of the
  acoustic gravitational wave power spectrum from a first order phase
  transition}},  {\em Phys. Rev. D} {\bf 96} (2017), no.~10 103520,
  [\href{http://arxiv.org/abs/1704.05871}{{\tt arXiv:1704.05871}}]. [Erratum:
  Phys.Rev.D 101, 089902 (2020)].

\bibitem{Kosowsky:2001xp}
A.~Kosowsky, A.~Mack, and T.~Kahniashvili, {\it {Gravitational radiation from
  cosmological turbulence}},  {\em Phys. Rev. D} {\bf 66} (2002) 024030,
  [\href{http://arxiv.org/abs/astro-ph/0111483}{{\tt astro-ph/0111483}}].

\bibitem{Caprini:2006jb}
C.~Caprini and R.~Durrer, {\it {Gravitational waves from stochastic
  relativistic sources: Primordial turbulence and magnetic fields}},  {\em
  Phys. Rev. D} {\bf 74} (2006) 063521,
  [\href{http://arxiv.org/abs/astro-ph/0603476}{{\tt astro-ph/0603476}}].

\bibitem{Gogoberidze:2007an}
G.~Gogoberidze, T.~Kahniashvili, and A.~Kosowsky, {\it {The Spectrum of
  Gravitational Radiation from Primordial Turbulence}},  {\em Phys. Rev. D}
  {\bf 76} (2007) 083002, [\href{http://arxiv.org/abs/0705.1733}{{\tt
  arXiv:0705.1733}}].

\bibitem{Caprini:2009yp}
C.~Caprini, R.~Durrer, and G.~Servant, {\it {The stochastic gravitational wave
  background from turbulence and magnetic fields generated by a first-order
  phase transition}},  {\em JCAP} {\bf 12} (2009) 024,
  [\href{http://arxiv.org/abs/0909.0622}{{\tt arXiv:0909.0622}}].

\bibitem{Niksa:2018ofa}
P.~Niksa, M.~Schlederer, and G.~Sigl, {\it {Gravitational Waves produced by
  Compressible MHD Turbulence from Cosmological Phase Transitions}},  {\em
  Class. Quant. Grav.} {\bf 35} (2018), no.~14 144001,
  [\href{http://arxiv.org/abs/1803.02271}{{\tt arXiv:1803.02271}}].

\bibitem{Caprini:2015zlo}
C.~Caprini et~al., {\it {Science with the space-based interferometer eLISA. II:
  Gravitational waves from cosmological phase transitions}},  {\em JCAP} {\bf
  04} (2016) 001, [\href{http://arxiv.org/abs/1512.06239}{{\tt
  arXiv:1512.06239}}].

\bibitem{Guo:2020grp}
H.-K. Guo, K.~Sinha, D.~Vagie, and G.~White, {\it {Phase Transitions in an
  Expanding Universe: Stochastic Gravitational Waves in Standard and
  Non-Standard Histories}},  {\em JCAP} {\bf 01} (2021) 001,
  [\href{http://arxiv.org/abs/2007.08537}{{\tt arXiv:2007.08537}}].

\bibitem{Hooper:2020evu}
D.~Hooper, G.~Krnjaic, J.~March-Russell, S.~D. McDermott, and
  R.~Petrossian-Byrne, {\it {Hot Gravitons and Gravitational Waves From Kerr
  Black Holes in the Early Universe}},
  \href{http://arxiv.org/abs/2004.00618}{{\tt arXiv:2004.00618}}.

\bibitem{Domenech:2021ztg}
G.~Dom\`enech, {\it {Scalar Induced Gravitational Waves Review}},  {\em
  Universe} {\bf 7} (2021), no.~11 398,
  [\href{http://arxiv.org/abs/2109.01398}{{\tt arXiv:2109.01398}}].

\bibitem{Papanikolaou:2022chm}
T.~Papanikolaou, {\it {Gravitational waves induced from primordial black hole
  fluctuations: the~effect of an extended mass function}},  {\em JCAP} {\bf 10}
  (2022) 089, [\href{http://arxiv.org/abs/2207.11041}{{\tt arXiv:2207.11041}}].

\bibitem{Borah:2022iym}
D.~Borah, S.~Jyoti~Das, R.~Samanta, and F.~R. Urban, {\it {PBH-infused seesaw
  origin of matter and unique gravitational waves}},  {\em JHEP} {\bf 03}
  (2023) 127, [\href{http://arxiv.org/abs/2211.15726}{{\tt arXiv:2211.15726}}].

\bibitem{Weltman:2018zrl}
A.~Weltman et~al., {\it {Fundamental physics with the Square Kilometre Array}},
   {\em Publ. Astron. Soc. Austral.} {\bf 37} (2020) e002,
  [\href{http://arxiv.org/abs/1810.02680}{{\tt arXiv:1810.02680}}].

\bibitem{Garcia-Bellido:2021zgu}
J.~Garcia-Bellido, H.~Murayama, and G.~White, {\it {Exploring the Early
  Universe with Gaia and THEIA}},  \href{http://arxiv.org/abs/2104.04778}{{\tt
  arXiv:2104.04778}}.

\bibitem{Sesana:2019vho}
A.~Sesana et~al., {\it {Unveiling the gravitational universe at $\mu$-Hz
  frequencies}},  {\em Exper. Astron.} {\bf 51} (2021), no.~3 1333--1383,
  [\href{http://arxiv.org/abs/1908.11391}{{\tt arXiv:1908.11391}}].

\bibitem{2017arXiv170200786A}
{\bf LISA} Collaboration, P.~Amaro-Seoane~et al, {\it {Laser Interferometer
  Space Antenna}},  {\em arXiv e-prints} (Feb., 2017) arXiv:1702.00786,
  [\href{http://arxiv.org/abs/1702.00786}{{\tt arXiv:1702.00786}}].

\bibitem{AEDGE:2019nxb}
{\bf AEDGE} Collaboration, Y.~A. El-Neaj et~al., {\it {AEDGE: Atomic Experiment
  for Dark Matter and Gravity Exploration in Space}},  {\em EPJ Quant.
  Technol.} {\bf 7} (2020) 6, [\href{http://arxiv.org/abs/1908.00802}{{\tt
  arXiv:1908.00802}}].

\bibitem{LIGOScientific:2014pky}
{\bf LIGO Scientific} Collaboration, J.~Aasi et~al., {\it {Advanced LIGO}},
  {\em Class. Quant. Grav.} {\bf 32} (2015) 074001,
  [\href{http://arxiv.org/abs/1411.4547}{{\tt arXiv:1411.4547}}].

\bibitem{Samanta:2021mdm}
R.~Samanta and F.~R. Urban, {\it {Testing Super Heavy Dark Matter from
  Primordial Black Holes with Gravitational Waves}},
  \href{http://arxiv.org/abs/2112.04836}{{\tt arXiv:2112.04836}}.

\bibitem{Baker:2019ndr}
M.~J. Baker, J.~Kopp, and A.~J. Long, {\it {Filtered Dark Matter at a First
  Order Phase Transition}},  {\em Phys. Rev. Lett.} {\bf 125} (2020), no.~15
  151102, [\href{http://arxiv.org/abs/1912.02830}{{\tt arXiv:1912.02830}}].

\bibitem{Domenech:2023mqk}
G.~Dom\`enech and M.~Sasaki, {\it {Gravitational wave hints black hole remnants
  as dark matter}},  {\em Class. Quant. Grav.} {\bf 40} (2023), no.~17 177001,
  [\href{http://arxiv.org/abs/2303.07661}{{\tt arXiv:2303.07661}}].

\bibitem{Masina:2020xhk}
I.~Masina, {\it {Dark matter and dark radiation from evaporating primordial
  black holes}},  {\em Eur. Phys. J. Plus} {\bf 135} (2020), no.~7 552,
  [\href{http://arxiv.org/abs/2004.04740}{{\tt arXiv:2004.04740}}].

\bibitem{Bernal:2020kse}
N.~Bernal and O.~Zapata, {\it {Self-interacting Dark Matter from Primordial
  Black Holes}},  {\em JCAP} {\bf 03} (2021) 007,
  [\href{http://arxiv.org/abs/2010.09725}{{\tt arXiv:2010.09725}}].

\bibitem{Irsic:2017ixq}
V.~Ir\v{s}i\v{c} et~al., {\it {New Constraints on the free-streaming of warm
  dark matter from intermediate and small scale Lyman-$\alpha$ forest data}},
  {\em Phys. Rev. D} {\bf 96} (2017), no.~2 023522,
  [\href{http://arxiv.org/abs/1702.01764}{{\tt arXiv:1702.01764}}].

\bibitem{Ballesteros:2020adh}
G.~Ballesteros, M.~A.~G. Garcia, and M.~Pierre, {\it {How warm are non-thermal
  relics? Lyman-$\alpha$ bounds on out-of-equilibrium dark matter}},  {\em
  JCAP} {\bf 03} (2021) 101, [\href{http://arxiv.org/abs/2011.13458}{{\tt
  arXiv:2011.13458}}].

\bibitem{DEramo:2020gpr}
F.~D'Eramo and A.~Lenoci, {\it {Lower Mass Bounds on FIMPs}},
  \href{http://arxiv.org/abs/2012.01446}{{\tt arXiv:2012.01446}}.

\bibitem{Diamanti:2017xfo}
R.~Diamanti, S.~Ando, S.~Gariazzo, O.~Mena, and C.~Weniger, {\it {Cold dark
  matter plus not-so-clumpy dark relics}},  {\em JCAP} {\bf 06} (2017) 008,
  [\href{http://arxiv.org/abs/1701.03128}{{\tt arXiv:1701.03128}}].

\bibitem{Musco:2004ak}
I.~Musco, J.~C. Miller, and L.~Rezzolla, {\it {Computations of primordial black
  hole formation}},  {\em Class. Quant. Grav.} {\bf 22} (2005) 1405--1424,
  [\href{http://arxiv.org/abs/gr-qc/0412063}{{\tt gr-qc/0412063}}].

\bibitem{Kawana:2022olo}
K.~Kawana, T.~Kim, and P.~Lu, {\it {PBH formation from overdensities in delayed
  vacuum transitions}},  {\em Phys. Rev. D} {\bf 108} (2023), no.~10 103531,
  [\href{http://arxiv.org/abs/2212.14037}{{\tt arXiv:2212.14037}}].

\bibitem{Hashino:2022tcs}
K.~Hashino, S.~Kanemura, T.~Takahashi, and M.~Tanaka, {\it {Probing first-order
  electroweak phase transition via primordial black holes in the effective
  field theory}},  {\em Phys. Lett. B} {\bf 838} (2023) 137688,
  [\href{http://arxiv.org/abs/2211.16225}{{\tt arXiv:2211.16225}}].

\bibitem{Jung:2021mku}
T.~H. Jung and T.~Okui, {\it {Primordial black holes from bubble collisions
  during a first-order phase transition}},
  \href{http://arxiv.org/abs/2110.04271}{{\tt arXiv:2110.04271}}.

\bibitem{Aggarwal:2020olq}
N.~Aggarwal et~al., {\it {Challenges and Opportunities of Gravitational Wave
  Searches at MHz to GHz Frequencies}},
  \href{http://arxiv.org/abs/2011.12414}{{\tt arXiv:2011.12414}}.

\bibitem{Chway:2019kft}
D.~Chway, T.~H. Jung, and C.~S. Shin, {\it {Dark matter filtering-out effect
  during a first-order phase transition}},  {\em Phys. Rev. D} {\bf 101}
  (2020), no.~9 095019, [\href{http://arxiv.org/abs/1912.04238}{{\tt
  arXiv:1912.04238}}].

\end{thebibliography}

\providecommand{\href}[2]{#2}\begingroup\raggedright\endgroup

\end{document}